\documentclass[%
 reprint,
preprintnumbers,
nofootinbib,
 amsmath,amssymb,
 aps,
]{revtex4-1}

\usepackage{graphicx}
\usepackage{dcolumn}
\usepackage{bm}

\begin{document}

\preprint{IPARCOS-UCM-23-036}

\title{Alleviation of anomalies from the non-oscillatory vacuum in loop quantum cosmology}

\author{Mercedes Mart\'in-Benito}
 \email{m.martin.benito@ucm.es}
\author{Rita B. Neves}%
 \email{rneves@ucm.es}
\affiliation{Departamento de F\'isica Te\'orica and IPARCOS, Universidad Complutense de Madrid, Parque de Ciencias 1, 28040 Madrid, Spain}

\author{Javier Olmedo}
 \email{javolmedo@ugr.es}
\affiliation{Departamento de F\'isica Te\'orica y del Cosmos, Universidad de Granada, Granada-18071, Spain}%

\date{\today}

\begin{abstract}
In this work we investigate observational signatures of a primordial power spectrum with exponential infrared suppression, motivated by the choice of a non-oscillatory vacuum in a bouncing and inflationary geometry within Loop Quantum Cosmology (LQC). We leave the parameter that defines the scale at which suppression occurs free and perform a Bayesian analysis, comparing with CMB data. The data shows a preference for some of the suppression to be within the observable window. Guided by this analysis, we choose concrete illustrative values for this parameter. We show that the model affects only slightly the parity anomaly, but it is capable of alleviating the lensing and power suppression anomalies.
\end{abstract}

\maketitle


\section{Introduction}

In the last decades, cosmology has reached a high degree of maturity as a research field thanks in part to the increasingly more accurate measurements of the Cosmic Microwave Background (CMB). For the most part, the observed CMB is well explained by the inflationary paradigm \cite{COBE}, where quantum fluctuations in the very early Universe seed the temperature fluctuations observed today. However, some anomalies have been identified in the data with respect to predictions from standard cosmology and persist in recent observations \cite{PlanckVII}. It is thought that these may be hints of non-standard processes occurring in the very early Universe. This has captured the attention of researchers in the field of quantum gravity, as it may open an observational window to the quantum nature of spacetime.

Within the approaches to quantum cosmology, Loop Quantum Cosmology (LQC) is one of the most promising ones in the literature \cite{LQCreview_Bojowald2005,LQCreview_Ashtekar2011,LQCreview_Banerjee2012,LQCreview_Agullo2016}. 
It applies the non-perturbative and background independent quantization program of Loop Quantum Gravity to cosmological models.
When applied to flat Friedmann-Lema\^itre-Robertson-Walker (FLRW) spacetimes minimally coupled to a scalar field that sources inflation, it generates non-trivial pre-inflationary dynamics that resolve the big-bang singularity in terms of a quantum bounce \cite{APS_PRL,APS_extended,Bentivegna2008,Kaminski2009,Pawlowski2012,Ashtekar2011}. This bounce occurs in a kinetically dominated epoch of the Universe and connects a contracting branch with an expanding one. Soon after the bounce, the Universe goes through a period of decelerated expansion, before the potential of the scalar field begins to dominate and standard slow-roll inflation begins. This affects the evolution of primordial perturbations, as some modes cross out of and back into the horizon before the onset of inflation. It is no longer reasonable to assume that they reach it in the Bunch-Davies vacuum of standard cosmology. This may then have an effect in the primordial power spectrum. In the standard $\Lambda$CDM model, this is a near-scale invariant power spectrum, with a slight red tilt. In the case of LQC, where several well motivated choices of vacua have been explored, the power spectrum is affected in the infrared, and the particular deviations from near-scale invariance depend on details of the quantization and on the vacuum choice (see \cite{Li2021,Agullo2023} for recent reviews and \cite{Barrau:2018gyz,Schander:2015eja,Han:2017wmt,Ashtekar2020,Ashtekar2021,Agullo2020,Agullo2021} for particular approaches).\footnote{Note that within the apporach of Refs. \cite{Barrau:2018gyz,Schander:2015eja} there are choices of initial conditions for perturbations that are ruled out since they do not yield a near-scale invariant scalar power spectrum in the ultraviolet.}

Previous works within LQC \cite{Ashtekar2020,Ashtekar2021,Agullo2020,Agullo2021} have shown that such departures from near scale invariance may alleviate some anomalies in observations of large scales. In \cite{Ashtekar2020,Ashtekar2021}, adopting the equations of motion of the so-called dressed metric approach \cite{Ashtekar:2009mb,dressedmetric_Agullo_CQG_2013} and a particular vacuum state for the cosmological perturbations \cite{Ashtekar2016}, it was shown that both the power suppression anomaly and the lensing anomaly may be alleviated. The first is related to a lack of power in the CMB for large multipoles, which is a consequence of the fact that the temperature-temperature correlation function is consistent with zero for large angular scales \cite{Copi2008,Copi2013,Copi2018}. This corresponds to a very unlikely realization of a $\Lambda$CDM universe, which predicts a much larger value of the corresponding estimator than what is observed. The effects of the LQC model considered in Refs. \cite{Ashtekar2020,Ashtekar2021} were able to alleviate the anomaly in the concrete sense that the expected value of this estimator is lower than in $\Lambda$CDM. However, one could argue that this alone is not enough to conclude an alleviation of the anomaly, and a computation of the distribution of the estimator would be necessary, so that the p-value of the observation may be found. The second anomaly was found as a consequence of a consistency test in $\Lambda$CDM. A phenomenological parameter is introduced to quantify how the CMB is lensed from the surface of last scattering until today. It is found to be incompatible with the prediction from $\Lambda$CDM at $\sim$2-$\sigma$ level. The LQC model of \cite{Ashtekar2020,Ashtekar2021} is able to alleviate this anomaly, by affecting the statistics of other parameters. It does not affect post-inflationary physics that relate to lensing, rather it shifts predictions enough that the inconsistencies no longer present as strongly as in the standard model. References \cite{Agullo2020,Agullo2021} consider primordial power spectra with power enhancement of different slopes.  These may be motivated by different models, of which LQC is an example. Here, non-Gaussianities become a key ingredient. They provide a mechanism that correlates the largest wavelength modes of the CMB and super-horizon (non-observable) modes. The effect is to modify the variance of the perturbations, even if the mean remains unaltered. Then, certain features become more likely in this scenario than in standard cosmology. This alleviates the aforementioned anomalies as well as the parity asymmetry anomaly, which refers to the fact that more power is observed in odd multipoles than in even ones, which is not predicted by $\Lambda$CDM.

In this work, we consider LQC but we depart from previous analyses in two ways, both related to the ambiguities present in the construction of the cosmological model at hand. First, we adopt the equations of motion derived from hybrid LQC \cite{Gomar2014,Gomar2015,hyb-vs-dress,hybrid:ElizagaNavascues2021} , mainly motivated by the fact that so far no Bayesian analysis comparing predictions with observations has been conducted adopting such prescription. One of its advantages is that the equations of motion for the perturbations are hyperbolic at the bounce, unlike those of the dressed metric approach \cite{hyb-vs-dress,Iteanu2022}.  The other main distinction from previous Bayesian analyses is the choice of initial conditions for perturbations, that we choose to be the so-called Non-Oscillatory (NO) vacuum.  This state has been motivated in previous LQC literature and shown to display interesting properties that make it an appealing vacuum choice \cite{deBlas2016,CastelloGomar2017,Elizaga_Navascu_s_2018, menava, ElizagaNavascues2021}. However a full understanding of its physical consequences requires a proper statistical analysis. This is the main aim of our present work, completing in this way those previous analyses based on this vacuum choice. More specifically, the NO vacuum minimizes oscillations in the primordial power spectrum, and leads to a power spectrum that is the near-scale invariant one of $\Lambda$CDM with exponential power suppression in the infrared and some small oscillations in intermediate scales. 
Furthermore, this state can be seen as a particular state of low energy, which minimizes the energy density when smeared along a given time window \cite{mno-sles}.
Vacua within the same family show also a strong suppression at infrared wavenumbers and free of oscillations \cite{menava}.
The scale at which these effects occur depends on initial conditions of the background at the bounce and the freedom in the choice of initial vacua there. Here, we leave this scale as a free parameter in a first instance. We perform a Bayesian analysis of the model, from which we are able to show that the data prefers some of the effects to be within the observable range. Guided by this analysis, we are able to fix some initial conditions and investigate their effect on the aforementioned anomalies. The goal is to investigate the observational consequences of this particular model with this choice of vacuum. On the other hand, this is also relevant for other vacua within LQC which lead to power suppression of infrared modes in the primordial power spectrum \cite{deBlas2016,Ashtekar2020,Agullo2020,Martin-Benito2021,ElizagaNavascues2021}. Thus, we contribute to the goal of understanding whether there are some robust features from LQC in predictions that transcend these ambiguities. Throughout we will perform comparisons with $\Lambda$CDM, by which we mean the standard inflationary paradigm with the Bunch-Davies vacuum for cosmological perturbations at the onset of inflation.

The structure of this papers is as follows. In section \ref{sec:anomalies} we briefly review the three aforementioned anomalies. Section \ref{sec:results in LQC} is dedicated to results for the model we are considering within LQC. We present the Bayesian analysis of the model with all parameters free, and then fix initial conditions to investigate possible alleviation of anomalies in concrete cases. Finally, section \ref{sec:conclusions} is dedicated to concluding remarks. We have also included two appendices with some details of our calculations.

\section{Anomalies in the CMB data}\label{sec:anomalies}

The temperature map of the CMB is remarkably uniform, with an average of $\bar{T} = 2.725 \pm 0.002 K$ \cite{Fixsen:1996nj} and fluctuations between different directions $\hat{n}$ of the order of $10^{-5} K$. These fluctuations can be expanded in spherical harmonics and described in terms of their coefficients $a_{\ell m}$:
\begin{equation}
    \delta T(\hat{n}) = \sum_{\ell m} a_{\ell m}\,Y_{\ell m}(\hat{n})
\end{equation}
Theoretical models, in particular $\Lambda$CDM, can only predict statistical properties of the CMB map. Therefore, we are particularly interested in the moments of the coefficients, rather than their actual values. Furthermore, the $\Lambda$CDM model predicts these fluctuations to be statistically isotropic and Gaussian, thus fully characterized by the mean and the second moment
\begin{equation}
    C(\theta) = \langle \delta T(\hat{n}), \delta T(\hat{n}^{\prime}) \rangle,
\end{equation}
where $\theta$ is the angle between two directions in the sky $\hat{n}$ and $\hat{n}^{\prime}$.
Homogeneity and isotropy imply that the second moments of $a_{\ell m}$ are diagonal and depend only on the multipole $\ell$ and can thus be fully characterized by the angular power-spectrum $C_{\ell}$:\footnote{Note that the angular power spectrum can always be defined as $\langle |a_{\ell m}|^2 \rangle$, but it is only in the case of random Gaussian statistically isotropic temperature fluctuations that it fully describes the second moment.}
\begin{equation}
    \langle a_{\ell m} a_{\ell' m'}^{\star}\rangle = C_{\ell} \delta_{\ell \ell'} \delta_{m m'}.
\end{equation}
The angular power-spectrum $C_{\ell}$ is related to the correlations in physical space $C(\theta)$ through:
\begin{equation}\label{eq:ctheta}
    C(\theta) = \frac{1}{4\pi} \sum_{\ell} (2\ell + 1) C_{\ell} P_{\ell}(\cos \theta),
\end{equation}
where $P_{\ell}(\cos \theta)$ are the Legendre polynomials. 

Early observations \cite{COBE} indicated that the fluctuations were consistent with an almost scale-invariant power spectrum. This is precisely the prediction of theoretical models within the inflationary paradigm, as long as inflation lasts long enough. Although the standard $\Lambda$CDM model's predictions are able to fit the data well in general, some anomalies have been identified and persist in more recent observations \cite{PlanckVII}. Let us briefly review three that are relevant in the context of quantum cosmology and bouncing scenarios.

\subsection{Power suppression anomaly}

Observations show that there is a lack of correlations at large angles (larger than $60^{\circ}$) with respect to the expected behavior for $\Lambda$CDM. This translates to some lack of power in the angular power spectrum for low $\ell$ and is thus commonly called the power suppression anomaly. However, what is most relevant is that the two-point correlation is remarkably consistent with zero for these scales, except for some anti-correlations at $180^{\circ}$. Numerically, the anomaly is best quantified via the estimator
\begin{equation}
    S_{1/2} = \int_{-1/2}^1 C^2(\theta) d(\cos \theta).
\end{equation}

To simplify calculations and avoid noise coming from $C(\theta)$, this quantity may be computed through $C_{\ell}$ by inserting \eqref{eq:ctheta} as
\begin{equation}\label{eq:S12fromCl}
    S_{1/2} = \sum_{\ell,\ell^{\prime} = 2}^{\ell_{\textrm{max}}} C_{\ell} I_{\ell \ell^{\prime}} C_{\ell^{\prime}},
\end{equation}
where $I_{\ell \ell^{\prime}}$ include the integrals of products of Legendre polynomials, and can be found in Appendix A of \cite{Copi2008}.\footnote{Note the difference in notation: the authors of that work refer to $\mathcal{I}_{\ell \ell^{\prime}}(x)$, which can be related to our notation through $I_{\ell \ell^{\prime}} = \frac{(2\ell+1)(2\ell^{\prime}+1)}{4\pi} \mathcal{I}_{\ell \ell^{\prime}}\left(x=1/2\right)$} The sum should in principle be over all (available) $\ell$, $\ell^{\prime}$, but it is enough to consider up to $\ell_{\textrm{max}} \sim 100$, as the Legendre polynomials sufficiently suppress higher multipole terms.

From Planck's cut-sky data (where the portion of the sky contaminated by our galactic disk has been removed through a mask), this quantity is found to be around $1200$, the exact value depending on the choice of map and mask \cite{PlanckVII}, whereas for full-sky data (where the contaminated region has been reconstructed) it is around $6700$ \cite{Copi2013}. These correspond to very unlikely realizations of the universe according to the $\Lambda$CDM model, where this quantity is expected to be around $35000$, with the observed values corresponding to p-values of $\sim 0.1 \%$ and $\sim 5\%$, respectively.\footnote{Here we define the p-values as the probability of finding values of $S_{1/2}$ at least as low as that observed in a random realization, given cosmic variance.} 

\subsection{Lensing anomaly}

As a consistency check, a phenomenological parameter, $A_L$, can be introduced to control how much or how little the CMB is lensed from the surface of last scattering until today, such that $A_L = 0$ means it is not at all lensed, and $A_L = 1$ corresponds to the prediction of the standard model. The anomaly is manifest in a Bayesian analysis of the $\Lambda$CDM + $A_L$ model, as it shows that the data prefers $A_L > 1$ at a 2-sigma level, and with large improvements of $\chi^2$.\footnote{The significance of this anomaly depends on the set of data. For the Planck 2018 data without lensing, this significance reaches about 3-$\sigma$, whereas for the data with lensing it is smaller than 2-$\sigma$. In this work we will consider the data with lensing, so that our results can be compared with those of other investigations in LQC that rely on the same data, such as \cite{Ashtekar2020,Ashtekar2021}.} If one also leaves the curvature of the Universe as a free parameter, then $A_L = 1$ is within the 1-sigma region if we allow the curvature to be negative, corresponding to a closed universe. However, this leads to discrepancies with other sets of data, which has been dubbed a ``crisis'' in cosmology \cite{DiValentino2019}. Concretely, the point of view adopted in \cite{DiValentino2019} is that fixing the curvature of the Universe to be flat hides the inconsistencies in the data that are observed when the curvature is left free. The parameter $A_L$ allows then to quantify these inconsistencies even when they are hidden behind the choice of flat Universe. Our viewpoint is that a resolution of this anomaly must address the overall inconsistencies between predictions and observations, without affecting the lensing physics of the CMB. 

\subsection{Parity anomaly}\label{sec:anomalies_parity}

The data also shows an anomalous power excess of odd-$\ell$ multipoles with respect to even ones for large angular scales ($\ell < 30$) in the angular correlation function $C_\ell$. Concretely, the parity asymmetry estimator is taken to be $R^{TT}(\ell_{\textrm{max}}) = D_+(\ell_{\textrm{max}})/D_-(\ell_{\textrm{max}})$, where
\begin{equation}
    D_{\pm}(\ell_{\textrm{max}}) = \frac{1}{\ell^{\pm}_{tot}} \sum_{\ell = 2,\ell_{\textrm{max}}}^{\pm} \frac{\ell(\ell+1)}{2\pi}C_{\ell}
\end{equation}
quantify the mean power contained in even (+) / odd (-) multipoles up to $\ell_{\textrm{max}}$, and $\ell^{\pm}_{tot}$ is the total number of even/ odd multipoles from 2 to $\ell_{\rm max}$. Although the $\Lambda$CDM model predicts neutral parity ($R^{TT} = 1$), it is found that $R^{TT}(\ell_{\textrm{max}}) < 1$ for low multipoles, with a statistical significance $\lesssim 2\sigma$ \cite{PlanckVII}. Previous studies have indicated this might be related with the power suppression anomaly \cite{Agullo2020,Agullo2021}.

\section{Loop Quantum Cosmology}\label{sec:results in LQC}

LQC is a non-perturbative and background independent approach based on the techniques of LQG, which are applied to cosmological models. Its main result is that of the resolution of the big-bang singularity in terms of a quantum bounce, which connects a contracting epoch of the Universe with an expanding one. It then opens a window for the study of pre-inflationary physics. In a flat FLRW Universe minimally coupled to a scalar field subject to a potential, this bounce is likely to take place in a kinetically dominated regime \cite{Ashtekar2011}, during which the Universe undergoes a short period of very rapid acceleration immediately after the bounce, followed by a period of decelerated expansion. Then, once the potential of the scalar field starts to dominate, standard slow-roll inflation begins. A brief review of the background dynamics in LQC can be found in Appendix \ref{sec:app_background}.

As mentioned in the previous section, a nearly scale-invariant primordial power spectrum of cosmological perturbations, such as the one obtained in inflationary models with enough inflation, reproduces remarkably well most of the features of the observed CMB. This limits any modification to this power spectrum, so as not to spoil the agreement with observations that is already achieved.

The pre-inflationary dynamics of LQC may affect the evolution of cosmological perturbations, and therefore the primordial power spectrum. In this setting it is not longer justified that perturbations reach the onset of inflation in the standard Bunch-Davies vacuum typically assumed in standard cosmology. Indeed, in LQC there is well defined dynamics before inflation, and one may instead choose to fix the vacuum at the bounce or even in the pre-bounce branch. Either way, the dynamics of the background will affect the evolution of at least some modes of the perturbations, which may then reach the onset of inflation in an excited state with respect to the Bunch-Davies vacuum. In turn, these modes will freeze-out during inflation in a different state than that of standard cosmology, and therefore lead to differences in the primordial power spectrum. Additionally, the equation of motion of scalar perturbations receives explicit quantum corrections, adding to the effect due to different background dynamics \cite{Li2021,Agullo2023}. These also depend on the choice of prescription to quantize the system with perturbations \cite{hyb-vs-dress}. In this work we consider the dynamics obtained through the hybrid LQC prescription \cite{Gomar2014,Gomar2015, hyb-vs-dress,hybrid:ElizagaNavascues2021}. Details of the computations including the equations of motion of the Fourier modes of the perturbations $k$ are provided in Appendix \ref{sec:app_EOMS}.

The question then becomes whether these departures from the near scale-invariant power spectrum of standard cosmology occur within the observable window. The concrete predictions of the primordial power spectrum depend on details of the procedure and, importantly, on the choice of vacuum. However, departures from near scale-invariance will occur always at infrared wavenumbers of the characteristic scale of the bounce given by $k_{\text{LQC}}$, which is related to the Ricci scalar (or equivalently the energy density) at the bounce. The value of $k_{\text{LQC}}$ also depends on the value of the inflaton field at the bounce, $\phi_B$. A larger $\phi_B$ will generate more e-folds of inflation, washing out the effects of the pre-inflationary dynamics on the power spectrum to more infrared scales, leading to lower $k_{\text{LQC}}$. Although some heuristic arguments may help fix this value, we will leave it as a parameter of our model throughout this work. In short, the pre-inflationary dynamics of LQC, or any bouncing model with a period of kinetic dominance prior to inflation, leads to a primordial power spectrum that agrees very well with the nearly scale-invariant one of standard cosmology for $k > k_{\text{LQC}}$, and that departs from it for $k \leq k_{\text{LQC}}$. The value of $k_{\text{LQC}}$ and how much the power spectrum differs from that of standard cosmology might depend not only on the precise details that define the LQC pre-inflationary dynamics of both the background and the cosmological perturbations, but also on the choice of vacuum state for the latter.

One of the choices of vacuum state that has received quite some attention in the LQC literature is the family of the so-called NO vacua \cite{deBlas2016,CastelloGomar2017,Elizaga_Navascu_s_2018,menava,ElizagaNavascues2021}. These states minimize the oscillations in the evolution of the power spectrum in a given time interval (here from the bounce to the onset of inflation). This indeed translates into a minimization of the oscillations in the power spectrum at the end of inflation as a function of the comoving wavenumbers $k$. Their motivation comes from the fact that a highly oscillatory behavior of the power spectrum can be understood as some vacuum excitations that could mask the information about the traces of the fundamental state of the inhomogeneities. In other words, this is a state for cosmological perturbations well adapted to the background dynamics from the bounce to the future. One would thus claim that the NO vacua are optimal to gain observational access to those regimes near the bounce where LQC effects are non-negligible. This is the point of view adopted in \cite{deBlas2016,CastelloGomar2017,Elizaga_Navascu_s_2018,menava,ElizagaNavascues2021}. 

\begin{figure}[t]
    \includegraphics[width=0.49\textwidth]{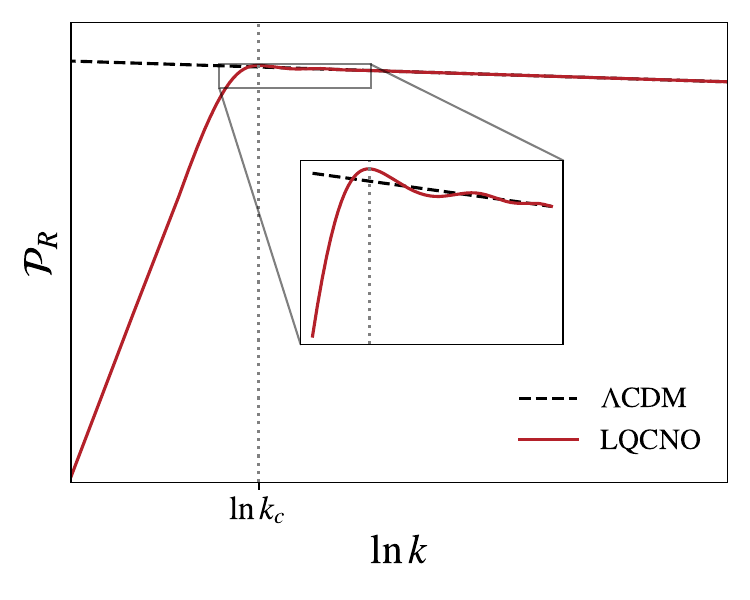}
    \caption{Typical primordial power spectrum of scalar perturbations for the LQCNO model (solid red line), as parametrized by \eqref{eq:PSNO}, and the corresponding power spectrum for the $\Lambda$CDM model (dashed black line).}
    \label{fig:PSNO}
\end{figure}

In this work, we will carry out a Bayesian analysis comparing Planck cosmological data with the physical predictions corresponding to choosing the NO vacuum of \cite{deBlas2016,CastelloGomar2017} as state of the perturbations in the context of LQC. We will refer to this model as LQCNO. Such an analysis, not done so far to the best of our knowledge, is essential to quantify how well the primordial power spectrum corresponding to the NO vacuum agrees with observations. Its power spectrum is exponentially suppressed for $k$ below a certain scale $k_c$, with some minimal oscillations for $k \gtrsim k_c$, as shown in Fig. \ref{fig:PSNO}. Note that this means that $k_{\text{LQC}}$ is somewhere ultraviolet of $k_c$, though the main modifications to the power spectrum occur infrared of it, via the power suppression. The two scales are proportional through a factor that depends on the value at the bounce of both the inflaton and the energy density (see the examples in Table \ref{tab:kcphiBefolds}). We will consider this scale to be fixed, as it was shown in \cite{Elizaga_Navascu_s_2018} that the primordial power spectrum arising from the NO vacuum is almost invariant under changes in that energy density scale. To simplify computations, we have parametrized this power spectrum with three free parameters, $k_c$, $A_s$ and $n_s$:
\begin{equation}\label{eq:PSNO}
    \mathcal{P}_{\mathcal{R}}(k) = f(k,k_c)\ \mathcal{P}_{\mathcal{R}}^{\Lambda\textrm{CDM}}(k),
\end{equation}
where
\begin{equation}
    \mathcal{P}_{\mathcal{R}}^{\Lambda\textrm{CDM}}(k) = A_s \left( \frac{k}{k_{\star}} \right)^{n_s-1}
\end{equation}
is the near scale-invariant power spectrum of the $\Lambda$CDM model, and $f(k,k_c)$ parametrizes the departure from it for LQCNO as described in Appendix \ref{sec:app_param}. Then the parameter $k_c$ encodes the freedom particular to the LQCNO model, which relate to the freedom in the choice of $\phi_B$, or equivalently the number of e-folds of inflation. For a more intuitive picture, throughout this work we will cast $k_c$ values into the corresponding (approximate) number of e-folds of inflation in models with quadratic and Starobinsky inflaton potentials. Approximate numerical expressions to relate these quantities can be found in Appendix \ref{sec:app_phiB_efolds} and \ref{sec:app_kcefolds}.

\begin{figure}[t]
    \includegraphics[width=0.49\textwidth]{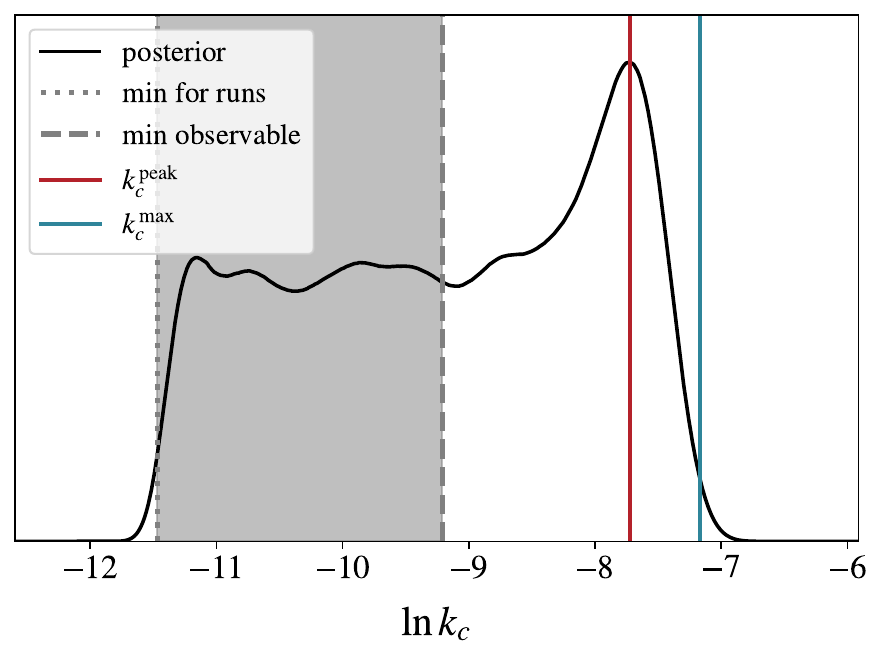}
    \caption{Marginalized posterior probability of the parameter $k_c$ of LQCNO resulting from a Bayesian analysis of the model with all its 7 parameters free, and integrating over the remaining 6. The grey region represents the portion of the explored parameter space that leads to a power spectrum where the modifications due to LQCNO are not observable.}
    \label{fig:posteriorkc}
\end{figure}

\subsection{Bayesian analysis}

Our model comprises of 7 free parameters: $k_c$, which encodes the freedom particular to LQCNO, and the 6 parameters of $\Lambda$CDM.\footnote{One could also consider the energy density scale of the bounce as an extra parameter. However, as explained before, the primordial power spectrum of the NO vacuum is insensitive to this scale, so we will not consider it as a free parameter in this case.} Of these, two refer to the parametrization of the primordial power spectrum, as mentioned in the previous section: $A_s$ and $n_s$. The remaining parameters are the baryonic and cold matter densities $\Omega_b h^2$ and $\Omega_c h^2$, the angular scale of acoustic oscillations $100 \theta_{\textrm{MC}}$, and the optical depth at reioinization $\tau_{\textrm{reio}}$. These four are relevant to characterize the post-inflationary Universe and model the propagation of perturbations from the surface of last scattering until today.

To compute predictions that can be compared with CMB observations, we have used the publicly available Boltzmann code \textit{CLASS} \cite{Blas_2011}, to which we have given externally the primordial power spectrum parametrized above. With this strategy there is no need for modifications of the code to accommodate LQC, as it is relevant only inasmuch as the primordial power spectrum is affected. The post-inflationary processes will not be modelled any differently with respect to standard cosmology and thus \textit{CLASS} can be used directly to simulate them. This will allow us to obtain the angular power spectrum $C_{\ell}$ for a given point in parameter space. Furthermore, we have resorted to \textit{MontePython} \cite{Brinckmann2018,Audren2012}, a sampler which is interfaced with \textit{CLASS}, to apply the Markov-Chain-Monte-Carlo (MCMC) method that we have used to explore the parameter space and perform the Bayesian analysis of this work. The analysis has been performed using the CMB \textit{Planck} 2018 data with lensing \cite{PlanckV,PlanckVIII}.

We will start by analysing the freedom in $k_c$. To do so, let us first clarify its physical meaning. As explained previously, this scale is closely related with $k_{\text{LQC}}$. Concretely, they are monotonously related and it is always true by construction that $k_{\text{LQC}} \geq k_c$. 
Then, if $k_c$ is sufficiently smaller than the minimum observable wavenumber, the signatures from LQCNO will not be visible, as the observed portion of the power spectrum will perfectly agree with the standard one. 
Conversely, if $k_c$ is larger than the minimum observed value, the corresponding power spectrum will be such that the power suppression with respect to near scale-invariance is within the observable window.

The posterior probability resulting from the Bayesian analysis is thus non-Gaussian in $k_c$ as it necessarily plateaus for low values, which correspond to a primordial power spectrum that agrees with that of $\Lambda$CDM in the observable range. The analysis of such a posterior needs to be performed carefully, as the typical parameters of best-fit (the value of the parameters at the peak of the distribution) and 1-$\sigma$ region (the region that encapsulates $68\%$ of the total volume of the posterior) do not hold the usual physical meaning in this case. Although we find that the best-fit of $k_c$ is below the observable range, without the context of the corresponding 1-$\sigma$ interval we have no physical interpretation for this information. In other words, since the probability distribution is not Gaussian, in our case the best fit artificially singles out a particular realization of the model that could be far from being a very likely one.

With this in mind, we focus our analysis on the marginalized posterior probability for $k_c$ shown in Fig. \ref{fig:posteriorkc}, obtained by analysing the LQCNO model with all 7 parameters free, and integrating the posterior over the other 6. This way, the marginalized posterior at each point in $k_c$ is informed by the whole 6-dimensional space of the remaining parameters. The cut-off of the posterior for low $k_c$ is not a reflection of a preference in the data, but rather of the minimum allowed for the runs. Then, the portion of the posterior probability corresponding to $k_c$ lower than the minimum observed is roughly a plateau, as expected. On the other hand, for scales larger than the minimum observed one, the  posterior probability displays a maximum followed by a very sharp cut-off. Let us now note that this cut-off is not a reflection of the maximum allowed for the runs, which was set at a much larger value. In this instance it does represent an actual preference in the data for $k_c$ to be below a certain scale. Nevertheless, there is a clear preference for some of the effects of LQCNO to be within the observable window, as the maximum likelihood corresponds to a value of $k_c$ that is observable. We will denote this as $k_c^{\textrm{peak}}$ in the following sections.

This result agrees qualitatively with that of \cite{Contaldi2003}, where a similar parametrization is used for the power spectrum and that served as inspiration for the parametrization used here. The difference lies essentially in the form of the suppression at infrared scales and in the amplitude of oscillations. Additionally, the motivation for the two cases are different, as in \cite{Contaldi2003} the power spectrum arises from a classical model with a period of kinetic dominance followed by a de Sitter branch, with only two free parameters: $n_s$ and a scale related to our $k_c$. 

In summary, the data prefers a $k_c$ that is observable over one that is not (which corresponds to an observable power spectrum equivalent to that of $\Lambda$CDM), but it very strongly constrains it. In other words the marginalized posterior probability of $k_c$ indicates that some departure from $\Lambda$CDM is preferred. This is further supported by the fact that both the minimum and average of the $\chi^2$ statistic are slightly lower for the LQCNO model with $k_c$ fixed at $k_c^{\rm peak}$ than that for $\Lambda$CDM (with $\Delta {\rm min}\chi^2 \simeq 1.7$, and $\Delta {\rm mean}\chi^2 \simeq 0.5$). This improvement is also discussed in Refs. \cite{Agullo2020,Agullo2021}. However, Refs. \cite{Ashtekar2020,Ashtekar2021} do not report this value.

In the rest of this analysis, we will consider two separate models. The first is LQCNO when fixing $k_c$ to $k_c^{\rm peak}$. The second is LQCNO when $k_c$ is fixed to $k_c^{\rm max}$, which we consider as the limit of maximum $k_c$, considering agreement with observations. The concrete values of these scales, as well as the corresponding values of $\phi_B$ and the number of e-folds of inflation $N$ are given in Table \ref{tab:kcphiBefolds} for quadratic and Starobinsky inflation. For comparison, Table \ref{tab:kcphiBefolds} also displays the corresponding value of the characteristic scale of LQC: $k_{\text{LQC}}\equiv a_\text{B}\sqrt{R_\text{B}/6}$, where $R$ is the scalar curvature and the subscript B denotes evaluation at the bounce.\footnote{In Appendix \ref{sec:app_scales} we give details on how to translate wave numbers $k$ expressed in natural units to ${\rm Mpc}^{-1}$.}

\begin{table*}[t]
    \centering
    \begin{tabular}{c|c|c|c|c|c|c}
         {} & \multicolumn{3}{c|}{quadratic} & \multicolumn{3}{c}{Starobinsky}\\
         $k_c ({\rm Mpc}^{-1})$ & $\phi_B$ & $N$ & $k_{\text LQC} ({\rm Mpc}^{-1})$ & $\phi_B$ & $N$ & $k_{\text LQC}({\rm Mpc}^{-1})$\\
         \hline
         $k_c^{\rm peak} = 4.44\times 10^{-4}$ & $0.940$ & $64.6$ & $0.635$ & $-1.460$ & $61.4$ & $1.12$ \\
         $k_c^{\rm max} = 7.70\times 10^{-4}$ & $0.925$ & $64.1$ & $2.03$ & $-1.462$ & $60.8$ & $3.41$
    \end{tabular}
    \caption{Values of the parameter $k_c$ used in this work cast into the corresponding values of $\phi_B$ and number of e-folds of inflation for the quadratic and Starobinsky inflation potentials. We also include the corresponding values of $k_{\text{LQC}}$. More details on these calculations can be found in appendices \ref{sec:app_scales},  \ref{sec:app_phiB_efolds} and \ref{sec:app_kcefolds}.}
    \label{tab:kcphiBefolds}
\end{table*}

\begin{figure*}[t]
    \centering
    \includegraphics[width=\textwidth]{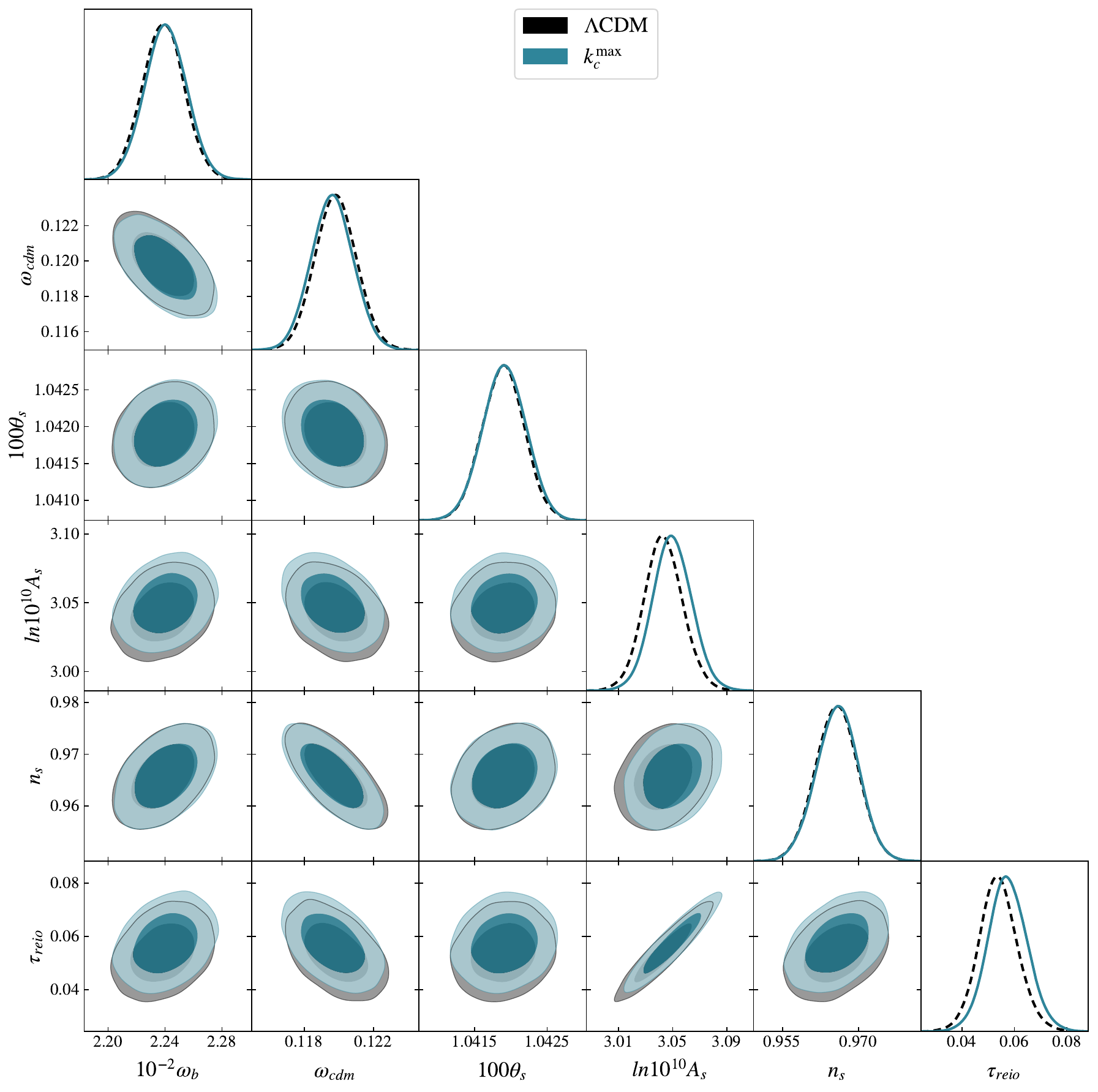}
    \caption{1 and 2-$\sigma$ C.L. 2D contours for the 6 $\Lambda$CDM parameters, plus 1D marginalized posteriors, for $\Lambda$CDM (black) and LQCNO with $k_c^{\rm max}$ (blue).}
    \label{fig:triangle}
\end{figure*}

Let us now compare the posterior probabilities of the 6 $\Lambda$CDM parameters in the $\Lambda$CDM and LQCNO models, as shown in Fig. \ref{fig:triangle}. For clarity, in these figures we do not present the case of LQCNO with $k_c^{\rm peak}$, as it sits between $\Lambda$CDM and $k_c^{\rm max}$. Most parameters seem to be mostly unaffected, except for $A_s$ and $\tau_{\textrm{reio}}$, as is evident from the 1-dimensional marginalized posterior distributions. The shifts in the contour plots of Fig. \ref{fig:triangle} are due to shifts in these two parameters. These are known to be correlated. A more opaque surface of last scattering corresponds to a lower optical depth $\tau$, by construction, and will result in perturbations reaching us with less power, and thus lower $A_s$. It seems natural then that the suppression of infrared modes of the LQCNO power spectrum will be compensated by higher power at the (already ultraviolet) pivot scale $A_s$. Consequently $\tau_{\textrm{reio}}$ also increases. Although this is the least constrained parameter of $\Lambda$CDM, this offers an opportunity for a falsifiable picture of LQCNO in the future, as independent measurements of $\tau_{\textrm{reio}}$ will help to constrain it further and therefore will allow to distinguish between LQCNO and $\Lambda$CDM \cite{Gupt2017}.

\subsection{Alleviation of anomalies}

To investigate the possible alleviation of the anomalies, in this section we will fix $k_c$ to the values of Table \ref{tab:kcphiBefolds} within the range where agreement with data is still achievable: $k_c^{\rm peak}$ and $k_c^{\rm max}$. The goal is to show how much LQC may contribute to the alleviation of anomalies depending on the choice of this parameter. The remaining 6 parameters will be left free and a Bayesian analysis is performed of the two models.

\subsubsection{Power suppression}

For each value of $k_c$, fixing the remaining parameters to the best-fit values obtained from the Bayesian analysis, we compute the corresponding $C_\ell$ using \textit{CLASS}. From these we are able to compute $S_{1/2}$ through \eqref{eq:S12fromCl}. This is the expected value of $S_{1/2}$ for our model, shown in Table \ref{tab:S12_pvalues}, which decreases substantially the higher the $k_c$. This is in agreement with what has been found in other scenarios within LQC \cite{Ashtekar2021,Agullo2021}.

However, this alone is not enough to infer that the anomaly has been alleviated. Indeed, our model might be capable of lowering the expected value of this quantity, with respect to $\Lambda$CDM, but it might also affect the variance of its distribution such that the observed value still represents a very unlikely realization of a universe according to the model. In other words, it is necessary to compute the p-value of the observed $S_{1/2}$ in the context of our model. To do so, we need to take into account that the previously obtained $C_\ell$'s are random variables with a Gaussian distribution with a variance given by the cosmic variance. Note that $S_{1/2}$ is a sum of products of $C_{\ell}$'s, and therefore its distribution is not Gaussian. It is more straightforward to obtain it numerically, through a Monte Carlo method.

\begin{figure}[t]
    \includegraphics[width=0.49\textwidth]{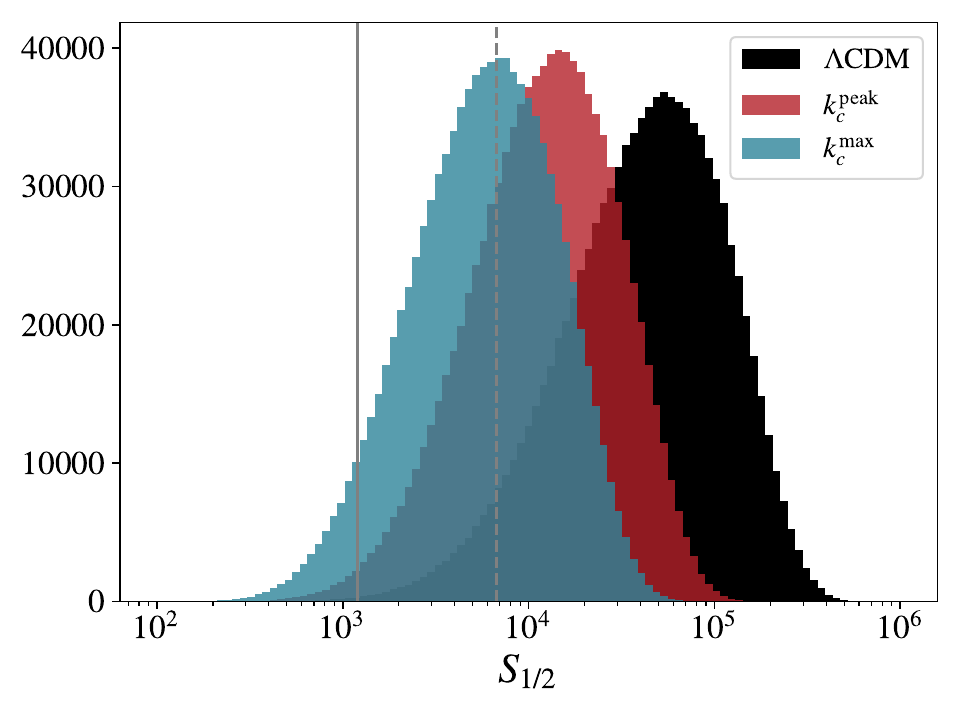}
    \caption{Distribution of $S_{1/2}$ considering a Gaussian distribution of cosmic variance for $C_{\ell}$'s. Also represented are the observed values for cut-sky (solid grey line) and full-sky data (dashed grey line).}
    \label{fig:S12}
\end{figure}

We have sampled the $C_{\ell}$ space randomly, computed $S_{1/2}$ for each point, and thus obtained the corresponding distributions of $S_{1/2}$, shown in Fig. \ref{fig:S12}. The p-value of the observed $S_{1/2}$ is simply the fraction of points that have resulted in $S_{1/2}$ at least as small as the observed one. As shown in Table \ref{tab:S12_pvalues}, the p-value improves substantially for higher $k_c$, both for the cut-sky and full-sky observations. In this manner we are able to say that the LQCNO model is capable of alleviating this anomaly, as the observed values are more likely realizations of the LQCNO model than of $\Lambda$CDM.
\begin{table}[t]
    \centering
    \begin{tabular}{c|c|c|c}
     {} & {} & \multicolumn{2}{c}{p-value} \\
     model & $S_{1/2}$ & cut-sky & full-sky\\
     \hline
     $\Lambda$CDM & 35430 &$\sim 0.1 \%$ & $\sim 5 \%$ \\
     LQCNO $k_c^{\text{peak}}$ & 14557 & $\sim 2 \%$ & $\sim 26 \%$ \\
     LQCNO $k_c^{\text{max}}$ & 7799 & $\sim 5 \%$ & $\sim 55 \%$ 
    \end{tabular}
    \caption{Expected values of $S_{1/2}$ and corresponding p-values with respect to observations for cut-sky ($S_{1/2} \sim 1200$) and full-sky data ($S_{1/2} \sim 6700$) for $\Lambda$CDM and LQCNO with different choices for $k_c$. Note that we define the p-value as the probability of obtaining a realization with $S_{1/2}$ at least as small as the observed one, according to the model.}
    \label{tab:S12_pvalues}
\end{table}
\begin{figure}[t]
    \includegraphics[width=0.49\textwidth]{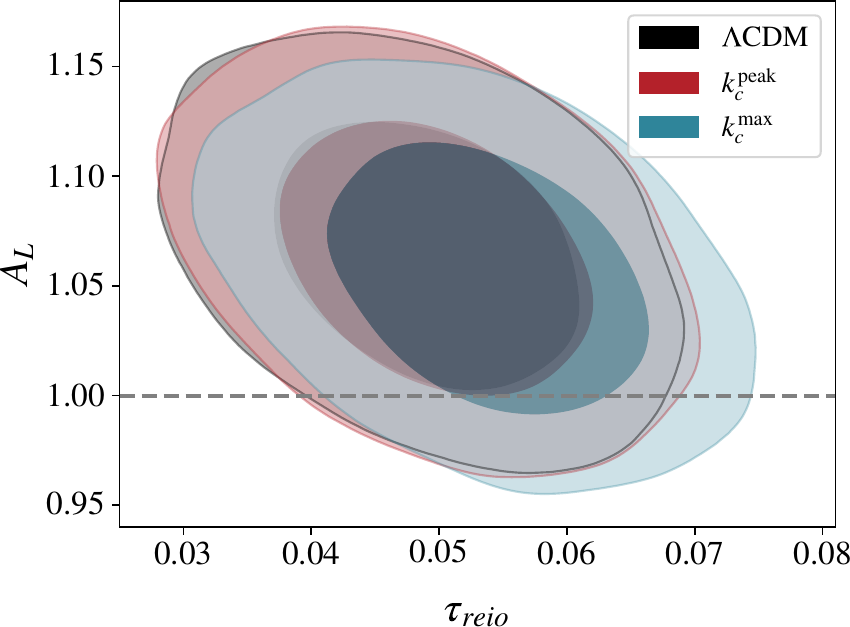}
    \caption{1 and 2-$\sigma$ C.L. 2D contour for the parameters $A_L$ and $\tau_{\textrm{reio}}$, for $\Lambda$CDM (black), LQCNO with $k_c = k_c^{\text{peak}}$ (red) and $k_c = k_c^{\text{max}}$ (blue).}
    \label{fig:AL}
\end{figure}

\subsubsection{Lensing anomaly}

For this analysis, we have included the extra parameter $A_L$ in our model, and performed a Bayesian analysis with the $\Lambda$CDM parameters as well as $A_L$ free. As mentioned, the best fit of almost all $\Lambda$CDM parameters is only sightly affected, except for $\tau_{\textrm{reio}}$ and $A_s$, which are correlated. Looking at the $A_L$ vs. $\tau$ contour plot of the posterior probability in Fig. \ref{fig:AL}, we can see that it shifts to higher values of $\tau_{\textrm{reio}}$ and lower values of $A_L$ as $k_c$ increases. For an appreciable $k_c$ within the observational window, this shift has pushed the 1-sigma region to include $A_L = 1$. Evidently, the dynamics of LQC does not affect the lensing the CMB goes through before it reaches our telescopes. Instead what we can conclude is that the model no longer presents the inconsistencies that are quantified through $A_L$ at the same level as the $\Lambda$CDM model.

Again, this analysis will benefit from future observations. As they constrain further and independently $\tau_{\textrm{reio}}$, they will help constrain $k_c$, and effectively constrain how much the dynamics of LQC may affect predictions in the observable window and contribute to the alleviation of anomalies.

\subsubsection{Parity anomaly}

It has been shown in \cite{Agullo2020,Agullo2021} that LQC may be able to alleviate the parity anomaly, when the power spectrum is such that non-Gaussianities become important. In those works, the considered power spectra show a power law suppression for infrared modes, as is the case in the LQCNO model, as well as some power enhancement for intermediate scales. Non-Gaussianities introduce a coupling between (non-observable) super-horizon modes and the largest observable ones. This affects the two-point correlation function, such that the mean value of the perturbations may remain unaltered, but the variance increases. In this sense, the anomalies are alleviated, as observations are more likely realizations in this scenario. Nevertheless, the concrete power spectra considered in those works can also affect the mean value of the parity asymmetry statistic.

\begin{figure}[t]
    \centering
    \includegraphics[width=0.49\textwidth]{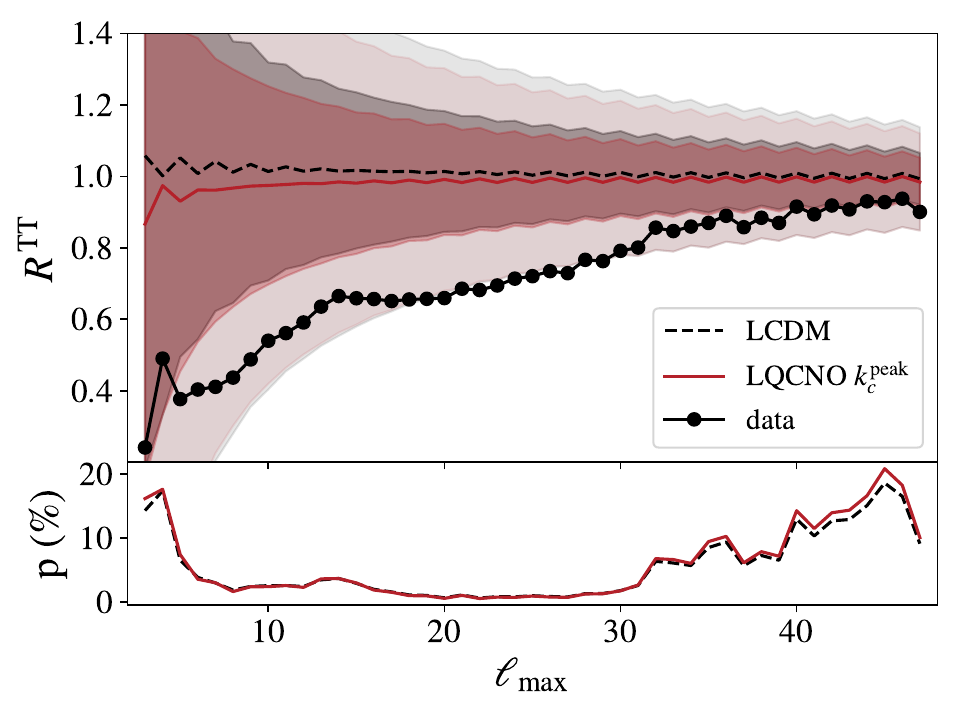}
    \caption{Upper: parity asymmetry statistic $R^{\rm TT}$ as a function of the maximum multipole $\ell_{\rm max}$ for $\Lambda$CDM (dashed black line) and LQCNO with $k_c^{\rm peak}$ (solid red line) and the corresponding Planck data (black dots). Also represented are the 1 and 2-$\sigma$ regions corresponding to $\Lambda$CDM (grey regions) and LQCNO with $k_c^{\rm max}$ (red regions). Lower: p-value of the observed data with respect to the models.}
    \label{fig:RTT_pvalue}
\end{figure}

\begin{figure}[t]
    \includegraphics[width=0.49\textwidth]{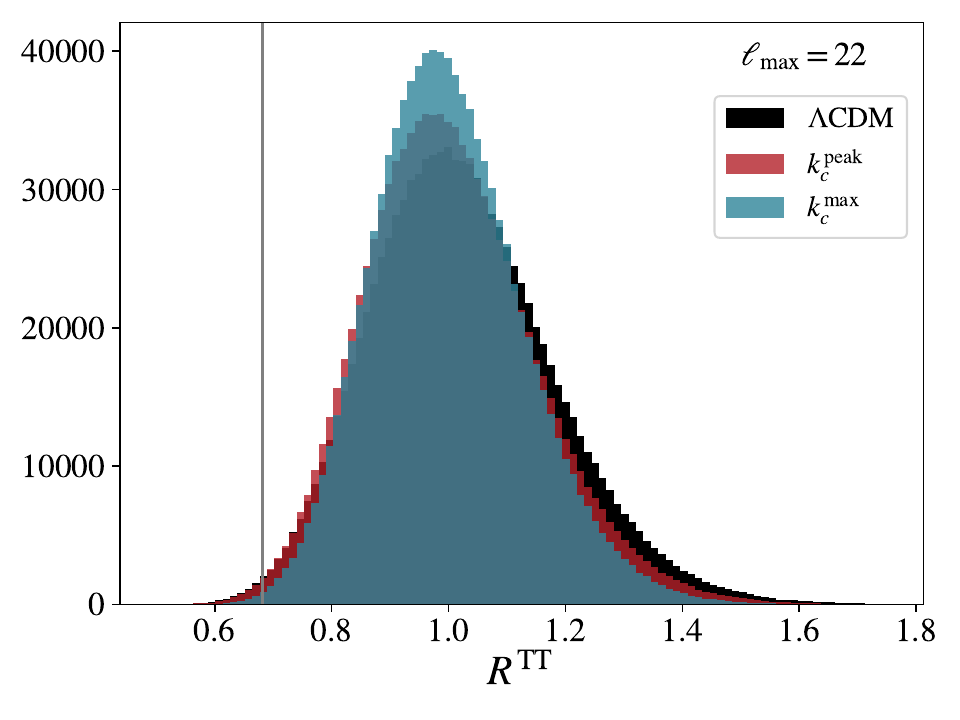}
    \caption{Distribution of $R^{\rm TT}$ for $\ell_{\rm max} = 22$ (lowest p-value) considering a Gaussian distribution of cosmic variance for $C_{\ell}$'s. Also represented is the corresponding observed value (solid grey line).}
    \label{fig:RTTdist}
\end{figure}

In this work we would like to understand the role of the power suppression of infrared modes in the alleviation of this anomaly. As such, we have computed the parity asymmetry statistic for the LQCNO model with $k_c^{\rm peak}$ as well as $\Lambda$CDM as outlined in section \ref{sec:anomalies_parity}, shown in the upper plot of Fig. \ref{fig:RTT_pvalue}. For clarity, we do not represent this statistic for LQCNO with $k_c^{\rm max}$, as it sits close to these two. It is not always below that of $k_c^{\rm peak}$, for some $\ell_{\rm max}$ it is between that of $k_c^{\rm peak}$ and $\Lambda$CDM. In other words, there does not seem to be a clear monotonic behavior of this quantity with $k_c$, as we have seen with $S_{1/2}$, $A_L$ and $\tau_{\rm reio}$. In any case, the LQCNO model introduces a small power asymmetry. However, as is the case of the power suppression anomaly, this is not enough to conclude an alleviation of the anomaly. Indeed, the p-values (represented in the lower plot) indicate that it is not the case for most $\ell_{\rm max}$. To compute them, we consider again the $C_{\ell}$ to be Gaussian random variables with cosmic variance, and sample them through a Monte Carlo method, obtaining the corresponding distributions of $R^{\rm TT}$ for each $\ell_{\rm max}$. In Fig. \ref{fig:RTTdist} we represent these distributions for the case of $\ell_{\rm max} = 22$, which results in the lowest p-value for all three models. Remarkably, from this figure it is clear that actually the distribution of $R^{\rm TT}$ becomes thinner for higher values of $k_c$ suppression. The slight shift in the mean is not sufficient to increase the p-value of the observation, which is in the tail of the distribution, and instead it actually decreases it for the particular case of $\ell_{\rm max} = 22$ represented in the figure. This is of course just one case, but it illustrates how the p-values may be lower even when the expected value of the estimator is closer to the data. Evidently for some $\ell_{\max}$ this is not the case and the p-value increases slightly in the case of LQCNO. Overall, the anomaly is essentially as strong as in the standard model. 

\section{Conclusions and discussion}\label{sec:conclusions}

In this work, we have aimed to perform a rigorous statistical analysis of the LQCNO model, via a comparison with Planck CMB data. On the one hand, the goal was to find possible signatures from this model in the data. One may argue that such predictions may also be obtained within the $\Lambda$CDM as long as one chooses the corresponding vacuum at the onset of inflation. In this case, such a choice would be ad-hoc, whereas in the LQC model that we have considered they are well motivated. In this spirit, we have first performed a Bayesian analysis with all 7 parameters  of our model free (6 of them coinciding with those of $\Lambda$CDM plus an extra scale $k_c$ inherent to our model). This allowed us to conclude that the data shows a preference for some of the effects particular to LQCNO to be within the observable window. This is quantified in the marginalized posterior probability of the parameter $k_c$, which shows a peak in the observable window, and a sharp cut-off shortly after. It seems that the LQCNO model may adjust the data as well as $\Lambda$CDM, and even slightly better when we fix the free parameter such that LQC effects are observable. We have also found that the LQCNO model with fixed $k_c$ affects appreciably the best-fit of the optical depth at reionization $\tau_{\rm reio}$. This opens the possibility of constraining $k_c$ with future observations of $\tau_{\rm reio}$. A more rigorous comparison between LQCNO and $\Lambda$CDM would require the computation of the Bayesian evidence. Given that this calls for computationally demanding methods to thoroughly explore the parameter space, we believe it is a tool that is most useful when we are able to constrain $k_c$ further from the data, so that we may compare $\Lambda$CDM with a model of LQCNO with fixed initial conditions motivated by observations. 

Furthermore, we have investigated the effect on anomalies of two illustrative models of LQCNO: one with initial conditions corresponding to the peak of the marginalized posterior distribution of $k_c$, and one in the tail of this distribution within the observational window, as a limiting case. We find that, for both the power suppression and the lensing anomalies, the more the effects are within the observational window the more the anomalies are alleviated. We have obtained the distribution of the estimator that quantifies the power suppression anomaly and found that the p-value of the observation improves in the LQCNO models with respect to $\Lambda$CDM. In the case of the lensing anomaly, we found that the 2D posterior probability of the lensing amplitude $A_L$ and optical depth at reionization is shifted, such that the 1-$\sigma$ region is pushed closer to $A_L = 1$. On the other hand, for the parity anomaly we have found that the power suppression that the LQCNO model offers in the primordial power spectrum is not enough to alleviate this tension, as it introduces only a slight asymmetry in the expected value, and in some cases the p-value of the data actually decreases. Additionally, the effect on this asymmetry does not seem to depend on the scale $k_c$ as much as the estimators of the previous anomalies. In this sense, given previous works in the literature \cite{Agullo2020,Agullo2021}, we believe non-Gaussianities would be relevant in this context and necessary for the alleviation of this anomaly.

On the other hand, this work is part of a larger effort to find robust results that are consequence of LQC in general, regardless of the ambiguities. We have shown that in the context of hybrid LQC, and with the particular choice of the NO vacuum, there is a preference for the effects of LQC to be observable and some alleviation of anomalies is possible. However, a power suppression of infrared modes is common in primordial power spectra of LQC models \cite{deBlas2016,Ashtekar2020,Agullo2020,Martin-Benito2021}. Any model that leads to a power spectrum with power suppression in the infrared will have the potential of alleviating these anomalies in the same way. More commonly, some power enhancement is also present for intermediate scales. The natural continuation of this work is to consider such power spectra and perform a similar analysis.

\acknowledgments

The authors thank G. A. Mena Marugán and V. Sreenath for valuable discussions. This work is supported by the Spanish Government through the projects PID2020-118159GB-C43, PID2020-118159GB-C44 and PID2019-105943GB-I00 (with FEDER contribution). R.~N. acknowledges financial support from Funda\c{c}\~ao para a Ci\^encia e a Tecnologia (FCT) through the research grant SFRH/BD/143525/2019. J. O. acknowledges the``Operative Program FEDER2014-2020 Junta de Andaluc\'ia-Consejer\'ia de Econom\'ia y Conocimiento" under project E-FQM-262-UGR18 by Universidad de Granada.

\appendix

\section{Background dynamics of LQC}\label{sec:app_background}

In this appendix we provide a brief summary of the effective equations of motion of  FLRW space-times in LQC. For more details about the physics of these models, see e.g. \cite{LQCreview_Ashtekar2011,APS_extended}. In LQC, the scale factor and Hubble parameter are quantum operators well defined on the states of the system, described by a wave function $\Psi$ on a suitable Hilbert space. The semi-classical sector of the theory is well represented by quantum states $\Psi$ that are sharply peaked on a classical geometry at late times (i.e. low curvatures). There,  general relativity is an excellent approximation. The expectation values of observables follow trajectories given by the so-called effective dynamics. They follow from an effective Hamiltonian that includes $\hbar$ corrections \cite{Taveras:2008ke, Diener:2014mia}. The phase space is a collection of pairs of canonical variables, the geometrical sector described by the Ashtekar-Barbero variables, and the matter sector by the scalar field and its momentum. The non-vanishing Poisson brackets are\footnote{Because the spatial slices are homogeneous, some integrals involved in the definition of the Hamiltonian and the symplectic form will diverge if the spatial slices are non-compact. This divergence is actually spurious. In order to regularize it, we restrict the integrals to a fiducial cell with large but finite coordinate volume $\mathcal{V}_0$. This volume can be taken to infinity at the end of the calculation. Hence, it acts as an infrared regulator. In any case, physical quantities will not depend on the value of $\mathcal{V}_0$.}.
\begin{equation}
\{c,p\}= \frac{8\pi G \gamma}{3{\mathcal{V}_0}}, \quad \left\{\phi, p_{\phi}\right\}=\frac{1}{\mathcal{V}_{0}}\, ,
\end{equation}
where $G$ is the Newton constant and $\gamma$ is the Barbero-Immirzi parameter. The variable $p$ is related to the scale factor $a$ of the space-time line element as $|p|= a^2$ and $c=\gamma\dot a/N_0$, being $N_0$ the lapse function. 

The leading order effective Hamiltonian takes the form
\begin{equation} \label{lqc-ham}
\mathcal{H}^{\rm LQC} = N_0 \mathcal{V}_0\bigg[  {\frac{-3|p|^{1/2}}{8\pi G\,\gamma^2\,}}  
\frac{\sin^2(\bar\mu\,c)}{\bar\mu^2} 
+\frac{p_{\phi}^2}{2 |p|^{3/2}}\, + |p|^{3/2}\, V(\phi)\bigg],
\end{equation}
where $\bar\mu \equiv\sqrt{\Delta/|p|}$ is the length of the edge of a (square) plaquette that encloses a physical area equal to $\Delta = 4 \sqrt{3} \pi \gamma G \hbar$, the smallest non-zero eigenvalue of the area operator in LQG. In the following, unless otherwise specified, we will choose $N_0=1$, which corresponds to cosmological time. 

Out of the Hamilton's equations, it is possible to derive the following modified Friedman equation:
\begin{equation}
H^2 = \frac{8 \pi G}{3} \rho \left( 1 - \frac{\rho}{\rho_c} \right).  
\end{equation}
Here $H=\dot a/a$ is the Hubble parameter and $\rho_c=3/8\pi G\,\gamma^2\Delta$ is the critical energy density. When $\rho=\rho_c$, the Hubble parameter vanishes. This corresponds to a cosmic bounce. Hence, $\rho_c$ determines the curvature scale at the bounce. But we should note that LQC effects are entirely negligible away from the bounce. In other words, significant departures from classical general relativity only happen for a very short time interval around the bounce of the order of the time $\sim \rho_c^{-1/4}$ in Planck units.

Furthermore, in order to evolve the equations of motion of the perturbations, we first need to specify the potential of the scalar field as well as initial conditions for background variables. We considered in this work the quadratic and Starobinsky type potentials:
\begin{equation}
    V_{\text{q}}(\phi) = \frac{1}{2}m^2 \phi^2,\qquad V_{\text{s}}(\phi)= V_0 \left(1-e^{-\sqrt{\frac{16\pi G}{3}}\phi}\right)^2,
\end{equation}
respectively. An agreement with observations requires $m = 1.21\times 10^{-6}$ and $V_0 = 1.77 \times 10^{-13}$ \cite{Bonga2015,Ashtekar2016}. For the initial data, in this model we only need to specify $\phi_B=\phi(t=t_B)$, since $H^2$ equals zero at the bounce, namely, at $t=t_B$. The kinetic energy of the scalar field is obtained from the condition $\rho(t=t_B)=\rho_c$. One should also specify the value of the scale factor at the bounce. We will discuss this below. 

\section{Details of numerical computations of perturbations}\label{sec:app_PPS}

\subsection{Equations of motion}\label{sec:app_EOMS}

In this work, we wish to obtain the primordial power spectrum of the comoving curvature perturbation ${\cal R}_k=u_{k}/z$, which is computed by
\begin{equation}
	\mathcal{P}_{\mathcal{R}}(k) = \frac{k^{3}}{2\pi^{2}}\frac{|u_{k}|^{2}}{z^{2}}\Big|_{\eta = \eta_{\rm end}}
\end{equation}
at conformal time $\eta_{\rm end}$ when inflation ends. Here, $z=a\dot{\phi}/H$, where the dot represents derivative with respect to cosmological time. Additionally $u_{k}$ obey the equations of motion
\begin{equation}\label{eq:EOMuk}
	u^{\prime\prime}_{\vec{k}}(\eta) + \left(k^2+s(\eta)\right) u_{\vec{k}}(\eta) = 0,
\end{equation}
where prime denotes derivative with respect to $\eta$, the so-called conformal time, defined as $d\eta = dt/a(t)$. The time-dependent mass term $s(\eta)$ depends on the particular model. For standard FLRW cosmologies $s(\eta) = -z^{\prime\prime}/z$. In the case of the hybrid LQC approach, 
\begin{equation}\label{eq:s(eta)}
\begin{split}
	s(\eta) =& -\frac{4\pi G}{3} a^2 \left(\rho - 3P\right) \\
 &+ a^2\left[V_{,\phi\phi}+48\pi G V(\phi)+ 6 \frac{a^{\prime}\phi^{\prime}}{a^3 \rho} V_{,\phi}-\frac{48\pi G}{\rho} V^2(\phi)\right].   
\end{split}
\end{equation}
It is dependent on background variables $a$ and $H$, $\rho$ (energy density of the inflaton) and $P$ (inflaton pressure), as well as on the inflaton potential $V(\phi)$ and derivatives of it with respect to $\phi$.

In general, there is no analytical solution to \eqref{eq:EOMuk} with time-dependent mass \eqref{eq:s(eta)}. It can be computed numerically given initial conditions $u_k(\eta_0)$, $u_k^{\prime}(\eta_0)$, the choice of which amounts to a choice of vacuum of the perturbations. As mentioned in the main body of this work, we choose it to be the NO vacuum of \cite{deBlas2016,CastelloGomar2017}.

As we mentioned, the initial conditions for the background are determined by the value of the inflaton at the bounce, $\phi_B$, the value of the scale factor at the bounce, and the critical value of the energy density (at which the bounce occurs). We will elaborate on the first two choices shortly. The third could in principle be left as a free parameter (which is equivalent to leaving $\gamma$ free), however the primordial power spectrum of the NO vacuum has been shown to be almost invariant under changes in the critical energy density. Thus, we fix $\gamma=0.2375$, as it is common in the LQC literature .

\subsection{Value of the scale factor at the bounce and relation between scales}\label{sec:app_scales}

As usual in the context of LQC, we also fix the scale factor to be 1 at the bounce, for convenience. However, in standard cosmology the scale factor is usually fixed to be 1 today. This choice is arbitrary, as the physical quantity is not the scale factor, but ratios of it. Nevertheless, to compare our predictions with observations we need to relate the two approaches, specifically taking into account that each one leads to different scales of the modes $k$. 

The procedure is as follows. We compute the dynamics of the background and of perturbations given our choice of scale factors, which results in a power spectrum $\mathcal{P}_{\mathcal{R}}(\tilde{k})$ for Fourier modes $\tilde{k}$. Given our choice of vacuum, this power spectrum is the near-scale invariant one of standard cosmology, with some oscillations and an exponential power suppression for infrared modes. The scale at which these departures from near-scale invariance occur depends on $\phi_B$. Then, to relate the scale $\tilde{k}$ with the one of standard cosmology, we resort to the pivot scale $k_{\star}$ as a reference scale. By definition, this is the scale at which the primordial power spectrum is $A_s$. For Planck data, $k_{\star} = 0.05\,{\rm Mpc}^{-1}$. Then, for a given $A_s$ we find the pivot scale in our units $\tilde{k}_{\star}$ as $\mathcal{P}_{\mathcal{R}}(\tilde{k}_{\star}) = A_s$. Finally, we can rescale $\tilde{k}$ to $k = \tilde{k}\cdot k_{\star}/\tilde{k}_{\star}$. The power spectrum in this scale may now be compared with observations from the Planck collaboration. The effect is to displace $\mathcal{P}_{\mathcal{R}}(\tilde{k})$ logarithmically in $\tilde{k}$.

\subsection{Parametrization of NO vacuum scalar primordial power spectrum}\label{sec:app_param}

In summary, the departures from near-scale invariant power spectrum occur at a scale that depends on both $\phi_B$ and $A_s$. Thus, to simplify calculations, we parametrize it with one free parameter that defines the scale at which exponential suppression occurs, encapsulating all the freedom of the power spectrum in our model. 

Inspired by \cite{Contaldi2003}, we have parametrized it as:
\begin{equation}
    f(k,k_c) = \begin{cases}
    N f_{\textrm{sup}}(k) f_H(\frac{k_d}{2}) & \textrm{if }k < \frac{k_d}{2},\\
    N f_{\textrm{sup}}(k) f_H(k) & \textrm{if } k \geq \frac{k_d}{2},
    \end{cases}
\end{equation}
where $f_{\textrm{sup}}(k) = 1- \exp(-(k/k_d)^{\lambda_c})$ parametrizes the exponential suppression, and 
\begin{align}
    f_H(k) =\frac{\pi}{2 k_d k}\Bigg|& k H_1^{(2)}\left(\frac{k}{k_d}\right)\sin\left(C\frac{k}{k_d}\right)+ \nonumber\\
    &\frac{k_d}{2}H_0^{(2)}\left(\frac{k}{k_d}\right)\bigg[2\frac{k}{k_d}\cos\left(C\frac{k}{k_d}\right)\nonumber\\
    &\phantom{\frac{k_d}{2}H_0^{(2)}\left(\frac{k}{k_d}\right)\bigg[}-\sin\left(C\frac{k}{k_d}\right)\bigg]\Bigg|^2,
\end{align}
where $H_n^{(2)}(x)$ is a specific Hankel function of the second kind, and $k_d \simeq k_c/1.7$. Here, $k_d$ separates the regions of oscillation and exponential suppression, and is the parameter that encodes the ambiguities coming from LQC, $N$ is a normalization factor so that $A_s$ maintains its meaning as in standard cosmology as the amplitude of the power spectrum at the pivot scale $k_{\star}$, therefore fixed at $N = (f_{\textrm{sup}}(k_{\star}) f_H(k_{\star}))^{-1}$, $\lambda_c = 2.95$ is the slope of the suppression, and $C = 2/1.8$ parametrizes the oscillations.

It is worth commenting that this parametrization is an ad hoc choice that fits well our NO power spectrum, but only for practical purposes, namely, for the simulations we carry out in $CLASS$ and the subsequent Bayesian analysis. However, we must remember that this parametrization has important limitations. For instance, it would correspond to a state that is not of Hadamard type, unlike the NO vacua.

\subsection{Relation between $\phi_B$ and number of e-folds}\label{sec:app_phiB_efolds}

As mentioned in the text, we leave the value of the inflaton field at the bounce, $\phi_B$, as a free parameter. This value affects how much inflation occurs. Consequently, the range of values explored is such that the resulting background dynamics produces enough e-folds for an agreement with observations to be possible. This depends heavily on the inflaton potential.

We find numerically that the number of e-folds of inflation $N$ is related to $\phi_B$ through:
\begin{equation}
    N = A \phi_B + B,
\end{equation}
where for a quadratic potential $A \simeq 38$ and $B \simeq 29$, whereas for a starobinsky potential $A \simeq 259$ and $B \simeq 440$. The number of e-folds from the bounce until inflation is approximately constant with $k_c$, around $4.4$ e-folds for the quadratic model and $4.9$ e-folds for the Starobinsky potential.

\subsection{Relation between $k_c$ and $\phi_B$}\label{sec:app_kcefolds}

In order to relate $k_c$ of the parametrization back to $\phi_B$ we provide in this section approximate relations obtained numerically considering $A_s$ to be fixed at the $\Lambda$CDM best-fit value.
We have found numerically
\begin{equation}
	\phi_B = C \ln k_c + D,
\end{equation}
where for the quadratic potential $C = -0.027$ and $D = 0.735$, whereas for the Starobinsky potential $C = -0.004$ and $D = -1.490$.

We consider this approximation to suffice, as the variation of $A_s$ at 1-$\sigma$ we have obtained from the Bayesian analysis is of $\sim 1.7 \%$. Considering a spectral index $n_s \simeq 0.96$ this impacts the shift in the scales $\tilde{k}$ explained in \ref{sec:app_scales} in $\sim 4 \%$, which would induce a variation of $< 0.1 \%$, according to the relations found above.

\bibliography{NOanomalies}

\begin{thebibliography}{51}%
\makeatletter
\providecommand \@ifxundefined [1]{%
 \@ifx{#1\undefined}
}%
\providecommand \@ifnum [1]{%
 \ifnum #1\expandafter \@firstoftwo
 \else \expandafter \@secondoftwo
 \fi
}%
\providecommand \@ifx [1]{%
 \ifx #1\expandafter \@firstoftwo
 \else \expandafter \@secondoftwo
 \fi
}%
\providecommand \natexlab [1]{#1}%
\providecommand \enquote  [1]{``#1''}%
\providecommand \bibnamefont  [1]{#1}%
\providecommand \bibfnamefont [1]{#1}%
\providecommand \citenamefont [1]{#1}%
\providecommand \href@noop [0]{\@secondoftwo}%
\providecommand \href [0]{\begingroup \@sanitize@url \@href}%
\providecommand \@href[1]{\@@startlink{#1}\@@href}%
\providecommand \@@href[1]{\endgroup#1\@@endlink}%
\providecommand \@sanitize@url [0]{\catcode `\\12\catcode `\$12\catcode
  `\&12\catcode `\#12\catcode `\^12\catcode `\_12\catcode `\%12\relax}%
\providecommand \@@startlink[1]{}%
\providecommand \@@endlink[0]{}%
\providecommand \url  [0]{\begingroup\@sanitize@url \@url }%
\providecommand \@url [1]{\endgroup\@href {#1}{\urlprefix }}%
\providecommand \urlprefix  [0]{URL }%
\providecommand \Eprint [0]{\href }%
\providecommand \doibase [0]{http://dx.doi.org/}%
\providecommand \selectlanguage [0]{\@gobble}%
\providecommand \bibinfo  [0]{\@secondoftwo}%
\providecommand \bibfield  [0]{\@secondoftwo}%
\providecommand \translation [1]{[#1]}%
\providecommand \BibitemOpen [0]{}%
\providecommand \bibitemStop [0]{}%
\providecommand \bibitemNoStop [0]{.\EOS\space}%
\providecommand \EOS [0]{\spacefactor3000\relax}%
\providecommand \BibitemShut  [1]{\csname bibitem#1\endcsname}%
\let\auto@bib@innerbib\@empty
\bibitem [{\citenamefont {Hinshaw}\ \emph {et~al.}(1996)\citenamefont
  {Hinshaw}, \citenamefont {Banday}, \citenamefont {Bennett}, \citenamefont
  {Gorski}, \citenamefont {Kogut}, \citenamefont {Lineweaver}, \citenamefont
  {Smoot},\ and\ \citenamefont {Wright}}]{COBE}%
  \BibitemOpen
  \bibfield  {author} {\bibinfo {author} {\bibfnamefont {G.}~\bibnamefont
  {Hinshaw}}, \bibinfo {author} {\bibfnamefont {A.~J.}\ \bibnamefont {Banday}},
  \bibinfo {author} {\bibfnamefont {C.~L.}\ \bibnamefont {Bennett}}, \bibinfo
  {author} {\bibfnamefont {K.~M.}\ \bibnamefont {Gorski}}, \bibinfo {author}
  {\bibfnamefont {A.}~\bibnamefont {Kogut}}, \bibinfo {author} {\bibfnamefont
  {C.~H.}\ \bibnamefont {Lineweaver}}, \bibinfo {author} {\bibfnamefont
  {G.~F.}\ \bibnamefont {Smoot}}, \ and\ \bibinfo {author} {\bibfnamefont
  {E.~L.}\ \bibnamefont {Wright}},\ }\href {\doibase 10.1086/310076} {\bibfield
   {journal} {\bibinfo  {journal} {Astrophys. J. Lett.}\ }\textbf {\bibinfo
  {volume} {464}},\ \bibinfo {pages} {L25} (\bibinfo {year} {1996})},\ \Eprint
  {http://arxiv.org/abs/astro-ph/9601061} {arXiv:astro-ph/9601061} \BibitemShut
  {NoStop}%
\bibitem [{\citenamefont {Akrami}\ \emph {et~al.}(2020)\citenamefont {Akrami}
  \emph {et~al.}}]{PlanckVII}%
  \BibitemOpen
  \bibfield  {author} {\bibinfo {author} {\bibfnamefont {Y.}~\bibnamefont
  {Akrami}} \emph {et~al.} (\bibinfo {collaboration} {Planck}),\ }\href
  {\doibase 10.1051/0004-6361/201935201} {\bibfield  {journal} {\bibinfo
  {journal} {Astron. Astrophys.}\ }\textbf {\bibinfo {volume} {641}},\ \bibinfo
  {pages} {A7} (\bibinfo {year} {2020})},\ \Eprint
  {http://arxiv.org/abs/1906.02552} {arXiv:1906.02552 [astro-ph.CO]}
  \BibitemShut {NoStop}%
\bibitem [{\citenamefont {Bojowald}(2005)}]{LQCreview_Bojowald2005}%
  \BibitemOpen
  \bibfield  {author} {\bibinfo {author} {\bibfnamefont {M.}~\bibnamefont
  {Bojowald}},\ }\href@noop {} {\bibfield  {journal} {\bibinfo  {journal}
  {Living Rev. Rel.}\ }\textbf {\bibinfo {volume} {8}},\ \bibinfo {pages} {11}
  (\bibinfo {year} {2005})},\ \Eprint {http://arxiv.org/abs/gr-qc/0601085}
  {arXiv:gr-qc/0601085 [gr-qc]} \BibitemShut {NoStop}%
\bibitem [{\citenamefont {Ashtekar}\ and\ \citenamefont
  {Singh}(2011)}]{LQCreview_Ashtekar2011}%
  \BibitemOpen
  \bibfield  {author} {\bibinfo {author} {\bibfnamefont {A.}~\bibnamefont
  {Ashtekar}}\ and\ \bibinfo {author} {\bibfnamefont {P.}~\bibnamefont
  {Singh}},\ }\href@noop {} {\bibfield  {journal} {\bibinfo  {journal} {Class.
  Quant. Grav.}\ }\textbf {\bibinfo {volume} {28}},\ \bibinfo {pages} {213001}
  (\bibinfo {year} {2011})},\ \Eprint {http://arxiv.org/abs/1108.0893}
  {arXiv:1108.0893 [gr-qc]} \BibitemShut {NoStop}%
\bibitem [{\citenamefont {Banerjee}\ \emph {et~al.}(2012)\citenamefont
  {Banerjee}, \citenamefont {Calcagni},\ and\ \citenamefont
  {Martin-Benito}}]{LQCreview_Banerjee2012}%
  \BibitemOpen
  \bibfield  {author} {\bibinfo {author} {\bibfnamefont {K.}~\bibnamefont
  {Banerjee}}, \bibinfo {author} {\bibfnamefont {G.}~\bibnamefont {Calcagni}},
  \ and\ \bibinfo {author} {\bibfnamefont {M.}~\bibnamefont {Martin-Benito}},\
  }\href@noop {} {\bibfield  {journal} {\bibinfo  {journal} {SIGMA}\ }\textbf
  {\bibinfo {volume} {8}},\ \bibinfo {pages} {016} (\bibinfo {year} {2012})},\
  \Eprint {http://arxiv.org/abs/1109.6801} {arXiv:1109.6801 [gr-qc]}
  \BibitemShut {NoStop}%
\bibitem [{\citenamefont {Agullo}\ and\ \citenamefont
  {Singh}(2017)}]{LQCreview_Agullo2016}%
  \BibitemOpen
  \bibfield  {author} {\bibinfo {author} {\bibfnamefont {I.}~\bibnamefont
  {Agullo}}\ and\ \bibinfo {author} {\bibfnamefont {P.}~\bibnamefont {Singh}},\
  }in\ \href@noop {} {\emph {\bibinfo {booktitle} {Loop Quantum Gravity: The
  First 30 Years}}},\ \bibinfo {editor} {edited by\ \bibinfo {editor}
  {\bibfnamefont {A.}~\bibnamefont {Ashtekar}}\ and\ \bibinfo {editor}
  {\bibfnamefont {J.}~\bibnamefont {Pullin}}}\ (\bibinfo  {publisher} {WSP},\
  \bibinfo {year} {2017})\ pp.\ \bibinfo {pages} {183--240},\ \Eprint
  {http://arxiv.org/abs/1612.01236} {arXiv:1612.01236 [gr-qc]} \BibitemShut
  {NoStop}%
\bibitem [{\citenamefont {Ashtekar}\ \emph
  {et~al.}(2006{\natexlab{a}})\citenamefont {Ashtekar}, \citenamefont
  {Pawlowski},\ and\ \citenamefont {Singh}}]{APS_PRL}%
  \BibitemOpen
  \bibfield  {author} {\bibinfo {author} {\bibfnamefont {A.}~\bibnamefont
  {Ashtekar}}, \bibinfo {author} {\bibfnamefont {T.}~\bibnamefont {Pawlowski}},
  \ and\ \bibinfo {author} {\bibfnamefont {P.}~\bibnamefont {Singh}},\
  }\href@noop {} {\bibfield  {journal} {\bibinfo  {journal} {Phys. Rev. Lett.}\
  }\textbf {\bibinfo {volume} {96}},\ \bibinfo {pages} {141301} (\bibinfo
  {year} {2006}{\natexlab{a}})},\ \Eprint {http://arxiv.org/abs/gr-qc/0602086}
  {arXiv:gr-qc/0602086 [gr-qc]} \BibitemShut {NoStop}%
\bibitem [{\citenamefont {Ashtekar}\ \emph
  {et~al.}(2006{\natexlab{b}})\citenamefont {Ashtekar}, \citenamefont
  {Pawlowski},\ and\ \citenamefont {Singh}}]{APS_extended}%
  \BibitemOpen
  \bibfield  {author} {\bibinfo {author} {\bibfnamefont {A.}~\bibnamefont
  {Ashtekar}}, \bibinfo {author} {\bibfnamefont {T.}~\bibnamefont {Pawlowski}},
  \ and\ \bibinfo {author} {\bibfnamefont {P.}~\bibnamefont {Singh}},\
  }\href@noop {} {\bibfield  {journal} {\bibinfo  {journal} {Phys. Rev.}\
  }\textbf {\bibinfo {volume} {D74}},\ \bibinfo {pages} {084003} (\bibinfo
  {year} {2006}{\natexlab{b}})},\ \Eprint {http://arxiv.org/abs/gr-qc/0607039}
  {arXiv:gr-qc/0607039 [gr-qc]} \BibitemShut {NoStop}%
\bibitem [{\citenamefont {Bentivegna}\ and\ \citenamefont
  {Pawlowski}(2008)}]{Bentivegna2008}%
  \BibitemOpen
  \bibfield  {author} {\bibinfo {author} {\bibfnamefont {E.}~\bibnamefont
  {Bentivegna}}\ and\ \bibinfo {author} {\bibfnamefont {T.}~\bibnamefont
  {Pawlowski}},\ }\href@noop {} {\bibfield  {journal} {\bibinfo  {journal}
  {Phys. Rev.}\ }\textbf {\bibinfo {volume} {D77}},\ \bibinfo {pages} {124025}
  (\bibinfo {year} {2008})},\ \Eprint {http://arxiv.org/abs/0803.4446}
  {arXiv:0803.4446 [gr-qc]} \BibitemShut {NoStop}%
\bibitem [{\citenamefont {Kaminski}\ and\ \citenamefont
  {Pawlowski}(2010)}]{Kaminski2009}%
  \BibitemOpen
  \bibfield  {author} {\bibinfo {author} {\bibfnamefont {W.}~\bibnamefont
  {Kaminski}}\ and\ \bibinfo {author} {\bibfnamefont {T.}~\bibnamefont
  {Pawlowski}},\ }\href@noop {} {\bibfield  {journal} {\bibinfo  {journal}
  {Phys. Rev.}\ }\textbf {\bibinfo {volume} {D81}},\ \bibinfo {pages} {024014}
  (\bibinfo {year} {2010})},\ \Eprint {http://arxiv.org/abs/0912.0162}
  {arXiv:0912.0162 [gr-qc]} \BibitemShut {NoStop}%
\bibitem [{\citenamefont {Pawlowski}\ and\ \citenamefont
  {Ashtekar}(2012)}]{Pawlowski2012}%
  \BibitemOpen
  \bibfield  {author} {\bibinfo {author} {\bibfnamefont {T.}~\bibnamefont
  {Pawlowski}}\ and\ \bibinfo {author} {\bibfnamefont {A.}~\bibnamefont
  {Ashtekar}},\ }\href@noop {} {\bibfield  {journal} {\bibinfo  {journal}
  {Phys. Rev.}\ }\textbf {\bibinfo {volume} {D85}},\ \bibinfo {pages} {064001}
  (\bibinfo {year} {2012})},\ \Eprint {http://arxiv.org/abs/1112.0360}
  {arXiv:1112.0360 [gr-qc]} \BibitemShut {NoStop}%
\bibitem [{\citenamefont {Ashtekar}\ and\ \citenamefont
  {Sloan}(2011)}]{Ashtekar2011}%
  \BibitemOpen
  \bibfield  {author} {\bibinfo {author} {\bibfnamefont {A.}~\bibnamefont
  {Ashtekar}}\ and\ \bibinfo {author} {\bibfnamefont {D.}~\bibnamefont
  {Sloan}},\ }\href {\doibase 10.1007/s10714-011-1246-y} {\bibfield  {journal}
  {\bibinfo  {journal} {Gen. Rel. Grav.}\ }\textbf {\bibinfo {volume} {43}},\
  \bibinfo {pages} {3619} (\bibinfo {year} {2011})},\ \Eprint
  {http://arxiv.org/abs/1103.2475} {arXiv:1103.2475 [gr-qc]} \BibitemShut
  {NoStop}%
\bibitem [{\citenamefont {Li}\ \emph {et~al.}(2021)\citenamefont {Li},
  \citenamefont {Singh},\ and\ \citenamefont {Wang}}]{Li2021}%
  \BibitemOpen
  \bibfield  {author} {\bibinfo {author} {\bibfnamefont {B.-F.}\ \bibnamefont
  {Li}}, \bibinfo {author} {\bibfnamefont {P.}~\bibnamefont {Singh}}, \ and\
  \bibinfo {author} {\bibfnamefont {A.}~\bibnamefont {Wang}},\ }\href {\doibase
  10.3389/fspas.2021.701417} {\bibfield  {journal} {\bibinfo  {journal} {Front.
  Astron. Space Sci.}\ }\textbf {\bibinfo {volume} {8}},\ \bibinfo {pages}
  {701417} (\bibinfo {year} {2021})},\ \Eprint
  {http://arxiv.org/abs/2105.14067} {arXiv:2105.14067 [gr-qc]} \BibitemShut
  {NoStop}%
\bibitem [{\citenamefont {Agull\'o}\ \emph {et~al.}(2023)\citenamefont
  {Agull\'o}, \citenamefont {Wang},\ and\ \citenamefont
  {Wilson-Ewing}}]{Agullo2023}%
  \BibitemOpen
  \bibfield  {author} {\bibinfo {author} {\bibfnamefont {I.}~\bibnamefont
  {Agull\'o}}, \bibinfo {author} {\bibfnamefont {A.}~\bibnamefont {Wang}}, \
  and\ \bibinfo {author} {\bibfnamefont {E.}~\bibnamefont {Wilson-Ewing}},\
  }\href@noop {} {\  (\bibinfo {year} {2023})},\ \Eprint
  {http://arxiv.org/abs/2301.10215} {arXiv:2301.10215 [gr-qc]} \BibitemShut
  {NoStop}%
\bibitem [{\citenamefont {Barrau}\ \emph {et~al.}(2018)\citenamefont {Barrau},
  \citenamefont {Jamet}, \citenamefont {Martineau},\ and\ \citenamefont
  {Moulin}}]{Barrau:2018gyz}%
  \BibitemOpen
  \bibfield  {author} {\bibinfo {author} {\bibfnamefont {A.}~\bibnamefont
  {Barrau}}, \bibinfo {author} {\bibfnamefont {P.}~\bibnamefont {Jamet}},
  \bibinfo {author} {\bibfnamefont {K.}~\bibnamefont {Martineau}}, \ and\
  \bibinfo {author} {\bibfnamefont {F.}~\bibnamefont {Moulin}},\ }\href
  {\doibase 10.1103/PhysRevD.98.086003} {\bibfield  {journal} {\bibinfo
  {journal} {Phys. Rev. D}\ }\textbf {\bibinfo {volume} {98}},\ \bibinfo
  {pages} {086003} (\bibinfo {year} {2018})},\ \Eprint
  {http://arxiv.org/abs/1807.06047} {arXiv:1807.06047 [gr-qc]} \BibitemShut
  {NoStop}%
\bibitem [{\citenamefont {Schander}\ \emph {et~al.}(2016)\citenamefont
  {Schander}, \citenamefont {Barrau}, \citenamefont {Bolliet}, \citenamefont
  {Linsefors}, \citenamefont {Mielczarek},\ and\ \citenamefont
  {Grain}}]{Schander:2015eja}%
  \BibitemOpen
  \bibfield  {author} {\bibinfo {author} {\bibfnamefont {S.}~\bibnamefont
  {Schander}}, \bibinfo {author} {\bibfnamefont {A.}~\bibnamefont {Barrau}},
  \bibinfo {author} {\bibfnamefont {B.}~\bibnamefont {Bolliet}}, \bibinfo
  {author} {\bibfnamefont {L.}~\bibnamefont {Linsefors}}, \bibinfo {author}
  {\bibfnamefont {J.}~\bibnamefont {Mielczarek}}, \ and\ \bibinfo {author}
  {\bibfnamefont {J.}~\bibnamefont {Grain}},\ }\href {\doibase
  10.1103/PhysRevD.93.023531} {\bibfield  {journal} {\bibinfo  {journal} {Phys.
  Rev. D}\ }\textbf {\bibinfo {volume} {93}},\ \bibinfo {pages} {023531}
  (\bibinfo {year} {2016})},\ \Eprint {http://arxiv.org/abs/1508.06786}
  {arXiv:1508.06786 [gr-qc]} \BibitemShut {NoStop}%
\bibitem [{\citenamefont {Han}\ and\ \citenamefont {Liu}(2018)}]{Han:2017wmt}%
  \BibitemOpen
  \bibfield  {author} {\bibinfo {author} {\bibfnamefont {Y.}~\bibnamefont
  {Han}}\ and\ \bibinfo {author} {\bibfnamefont {M.}~\bibnamefont {Liu}},\
  }\href {\doibase 10.1088/1361-6382/aab671} {\bibfield  {journal} {\bibinfo
  {journal} {Class. Quant. Grav.}\ }\textbf {\bibinfo {volume} {35}},\ \bibinfo
  {pages} {105017} (\bibinfo {year} {2018})},\ \Eprint
  {http://arxiv.org/abs/1711.04991} {arXiv:1711.04991 [gr-qc]} \BibitemShut
  {NoStop}%
\bibitem [{\citenamefont {Ashtekar}\ \emph {et~al.}(2020)\citenamefont
  {Ashtekar}, \citenamefont {Gupt}, \citenamefont {Jeong},\ and\ \citenamefont
  {Sreenath}}]{Ashtekar2020}%
  \BibitemOpen
  \bibfield  {author} {\bibinfo {author} {\bibfnamefont {A.}~\bibnamefont
  {Ashtekar}}, \bibinfo {author} {\bibfnamefont {B.}~\bibnamefont {Gupt}},
  \bibinfo {author} {\bibfnamefont {D.}~\bibnamefont {Jeong}}, \ and\ \bibinfo
  {author} {\bibfnamefont {V.}~\bibnamefont {Sreenath}},\ }\href {\doibase
  10.1103/PhysRevLett.125.051302} {\bibfield  {journal} {\bibinfo  {journal}
  {Phys. Rev. Lett.}\ }\textbf {\bibinfo {volume} {125}},\ \bibinfo {pages}
  {051302} (\bibinfo {year} {2020})},\ \Eprint
  {http://arxiv.org/abs/2001.11689} {arXiv:2001.11689 [astro-ph.CO]}
  \BibitemShut {NoStop}%
\bibitem [{\citenamefont {Ashtekar}\ \emph {et~al.}(2021)\citenamefont
  {Ashtekar}, \citenamefont {Gupt},\ and\ \citenamefont
  {Sreenath}}]{Ashtekar2021}%
  \BibitemOpen
  \bibfield  {author} {\bibinfo {author} {\bibfnamefont {A.}~\bibnamefont
  {Ashtekar}}, \bibinfo {author} {\bibfnamefont {B.}~\bibnamefont {Gupt}}, \
  and\ \bibinfo {author} {\bibfnamefont {V.}~\bibnamefont {Sreenath}},\ }\href
  {\doibase 10.3389/fspas.2021.685288} {\bibfield  {journal} {\bibinfo
  {journal} {Front. Astron. Space Sci.}\ }\textbf {\bibinfo {volume} {8}},\
  \bibinfo {pages} {76} (\bibinfo {year} {2021})},\ \Eprint
  {http://arxiv.org/abs/2103.14568} {arXiv:2103.14568 [gr-qc]} \BibitemShut
  {NoStop}%
\bibitem [{\citenamefont {Agullo}\ \emph
  {et~al.}(2021{\natexlab{a}})\citenamefont {Agullo}, \citenamefont {Kranas},\
  and\ \citenamefont {Sreenath}}]{Agullo2020}%
  \BibitemOpen
  \bibfield  {author} {\bibinfo {author} {\bibfnamefont {I.}~\bibnamefont
  {Agullo}}, \bibinfo {author} {\bibfnamefont {D.}~\bibnamefont {Kranas}}, \
  and\ \bibinfo {author} {\bibfnamefont {V.}~\bibnamefont {Sreenath}},\ }\href
  {\doibase 10.1007/s10714-020-02778-9} {\bibfield  {journal} {\bibinfo
  {journal} {Gen. Rel. Grav.}\ }\textbf {\bibinfo {volume} {53}},\ \bibinfo
  {pages} {17} (\bibinfo {year} {2021}{\natexlab{a}})},\ \Eprint
  {http://arxiv.org/abs/2005.01796} {arXiv:2005.01796 [astro-ph.CO]}
  \BibitemShut {NoStop}%
\bibitem [{\citenamefont {Agullo}\ \emph
  {et~al.}(2021{\natexlab{b}})\citenamefont {Agullo}, \citenamefont {Kranas},\
  and\ \citenamefont {Sreenath}}]{Agullo2021}%
  \BibitemOpen
  \bibfield  {author} {\bibinfo {author} {\bibfnamefont {I.}~\bibnamefont
  {Agullo}}, \bibinfo {author} {\bibfnamefont {D.}~\bibnamefont {Kranas}}, \
  and\ \bibinfo {author} {\bibfnamefont {V.}~\bibnamefont {Sreenath}},\ }\href
  {\doibase 10.3389/fspas.2021.703845} {\bibfield  {journal} {\bibinfo
  {journal} {Front. Astron. Space Sci.}\ }\textbf {\bibinfo {volume} {8}},\
  \bibinfo {pages} {703845} (\bibinfo {year} {2021}{\natexlab{b}})},\ \Eprint
  {http://arxiv.org/abs/2105.12993} {arXiv:2105.12993 [gr-qc]} \BibitemShut
  {NoStop}%
\bibitem [{\citenamefont {Ashtekar}\ \emph {et~al.}(2009)\citenamefont
  {Ashtekar}, \citenamefont {Kaminski},\ and\ \citenamefont
  {Lewandowski}}]{Ashtekar:2009mb}%
  \BibitemOpen
  \bibfield  {author} {\bibinfo {author} {\bibfnamefont {A.}~\bibnamefont
  {Ashtekar}}, \bibinfo {author} {\bibfnamefont {W.}~\bibnamefont {Kaminski}},
  \ and\ \bibinfo {author} {\bibfnamefont {J.}~\bibnamefont {Lewandowski}},\
  }\href {\doibase 10.1103/PhysRevD.79.064030} {\bibfield  {journal} {\bibinfo
  {journal} {Phys. Rev. D}\ }\textbf {\bibinfo {volume} {79}},\ \bibinfo
  {pages} {064030} (\bibinfo {year} {2009})},\ \Eprint
  {http://arxiv.org/abs/0901.0933} {arXiv:0901.0933 [gr-qc]} \BibitemShut
  {NoStop}%
\bibitem [{\citenamefont {Agullo}\ \emph {et~al.}(2013)\citenamefont {Agullo},
  \citenamefont {Ashtekar},\ and\ \citenamefont
  {Nelson}}]{dressedmetric_Agullo_CQG_2013}%
  \BibitemOpen
  \bibfield  {author} {\bibinfo {author} {\bibfnamefont {I.}~\bibnamefont
  {Agullo}}, \bibinfo {author} {\bibfnamefont {A.}~\bibnamefont {Ashtekar}}, \
  and\ \bibinfo {author} {\bibfnamefont {W.}~\bibnamefont {Nelson}},\ }\href
  {\doibase 10.1088/0264-9381/30/8/085014} {\bibfield  {journal} {\bibinfo
  {journal} {Classical and Quantum Gravity}\ }\textbf {\bibinfo {volume}
  {30}},\ \bibinfo {pages} {085014} (\bibinfo {year} {2013})},\ \Eprint
  {http://arxiv.org/abs/1302.0254} {arXiv:1302.0254 [gr-qc]} \BibitemShut
  {NoStop}%
\bibitem [{\citenamefont {Ashtekar}\ and\ \citenamefont
  {Gupt}(2017)}]{Ashtekar2016}%
  \BibitemOpen
  \bibfield  {author} {\bibinfo {author} {\bibfnamefont {A.}~\bibnamefont
  {Ashtekar}}\ and\ \bibinfo {author} {\bibfnamefont {B.}~\bibnamefont
  {Gupt}},\ }\href {\doibase 10.1088/1361-6382/aa52d4} {\bibfield  {journal}
  {\bibinfo  {journal} {Class. Quant. Grav.}\ }\textbf {\bibinfo {volume}
  {34}},\ \bibinfo {pages} {035004} (\bibinfo {year} {2017})},\ \Eprint
  {http://arxiv.org/abs/1610.09424} {arXiv:1610.09424 [gr-qc]} \BibitemShut
  {NoStop}%
\bibitem [{\citenamefont {Copi}\ \emph {et~al.}(2009)\citenamefont {Copi},
  \citenamefont {Huterer}, \citenamefont {Schwarz},\ and\ \citenamefont
  {Starkman}}]{Copi2008}%
  \BibitemOpen
  \bibfield  {author} {\bibinfo {author} {\bibfnamefont {C.~J.}\ \bibnamefont
  {Copi}}, \bibinfo {author} {\bibfnamefont {D.}~\bibnamefont {Huterer}},
  \bibinfo {author} {\bibfnamefont {D.~J.}\ \bibnamefont {Schwarz}}, \ and\
  \bibinfo {author} {\bibfnamefont {G.~D.}\ \bibnamefont {Starkman}},\ }\href
  {\doibase 10.1111/j.1365-2966.2009.15270.x} {\bibfield  {journal} {\bibinfo
  {journal} {Mon. Not. Roy. Astron. Soc.}\ }\textbf {\bibinfo {volume} {399}},\
  \bibinfo {pages} {295} (\bibinfo {year} {2009})},\ \Eprint
  {http://arxiv.org/abs/0808.3767} {arXiv:0808.3767 [astro-ph]} \BibitemShut
  {NoStop}%
\bibitem [{\citenamefont {Copi}\ \emph {et~al.}(2015)\citenamefont {Copi},
  \citenamefont {Huterer}, \citenamefont {Schwarz},\ and\ \citenamefont
  {Starkman}}]{Copi2013}%
  \BibitemOpen
  \bibfield  {author} {\bibinfo {author} {\bibfnamefont {C.~J.}\ \bibnamefont
  {Copi}}, \bibinfo {author} {\bibfnamefont {D.}~\bibnamefont {Huterer}},
  \bibinfo {author} {\bibfnamefont {D.~J.}\ \bibnamefont {Schwarz}}, \ and\
  \bibinfo {author} {\bibfnamefont {G.~D.}\ \bibnamefont {Starkman}},\ }\href
  {\doibase 10.1093/mnras/stv1143} {\bibfield  {journal} {\bibinfo  {journal}
  {Mon. Not. Roy. Astron. Soc.}\ }\textbf {\bibinfo {volume} {451}},\ \bibinfo
  {pages} {2978} (\bibinfo {year} {2015})},\ \Eprint
  {http://arxiv.org/abs/1310.3831} {arXiv:1310.3831 [astro-ph.CO]} \BibitemShut
  {NoStop}%
\bibitem [{\citenamefont {Copi}\ \emph {et~al.}(2019)\citenamefont {Copi},
  \citenamefont {Gurian}, \citenamefont {Kosowsky}, \citenamefont {Starkman},\
  and\ \citenamefont {Zhang}}]{Copi2018}%
  \BibitemOpen
  \bibfield  {author} {\bibinfo {author} {\bibfnamefont {C.~J.}\ \bibnamefont
  {Copi}}, \bibinfo {author} {\bibfnamefont {J.}~\bibnamefont {Gurian}},
  \bibinfo {author} {\bibfnamefont {A.}~\bibnamefont {Kosowsky}}, \bibinfo
  {author} {\bibfnamefont {G.~D.}\ \bibnamefont {Starkman}}, \ and\ \bibinfo
  {author} {\bibfnamefont {H.}~\bibnamefont {Zhang}},\ }\href {\doibase
  10.1093/mnras/stz2962} {\bibfield  {journal} {\bibinfo  {journal} {Mon. Not.
  Roy. Astron. Soc.}\ }\textbf {\bibinfo {volume} {490}},\ \bibinfo {pages}
  {5174} (\bibinfo {year} {2019})},\ \Eprint {http://arxiv.org/abs/1812.03946}
  {arXiv:1812.03946 [astro-ph.CO]} \BibitemShut {NoStop}%
\bibitem [{\citenamefont {Castell\'o~Gomar}\ \emph {et~al.}(2014)\citenamefont
  {Castell\'o~Gomar}, \citenamefont {Fern\'andez-M\'endez}, \citenamefont
  {Mena~Marug\'an},\ and\ \citenamefont {Olmedo}}]{Gomar2014}%
  \BibitemOpen
  \bibfield  {author} {\bibinfo {author} {\bibfnamefont {L.}~\bibnamefont
  {Castell\'o~Gomar}}, \bibinfo {author} {\bibfnamefont {M.}~\bibnamefont
  {Fern\'andez-M\'endez}}, \bibinfo {author} {\bibfnamefont {G.~A.}\
  \bibnamefont {Mena~Marug\'an}}, \ and\ \bibinfo {author} {\bibfnamefont
  {J.}~\bibnamefont {Olmedo}},\ }\href {\doibase 10.1103/PhysRevD.90.064015}
  {\bibfield  {journal} {\bibinfo  {journal} {Phys. Rev. D}\ }\textbf {\bibinfo
  {volume} {90}},\ \bibinfo {pages} {064015} (\bibinfo {year} {2014})},\
  \Eprint {http://arxiv.org/abs/1407.0998} {arXiv:1407.0998 [gr-qc]}
  \BibitemShut {NoStop}%
\bibitem [{\citenamefont {Castell\'o~Gomar}\ \emph {et~al.}(2015)\citenamefont
  {Castell\'o~Gomar}, \citenamefont {Mart\'\i{}n-Benito},\ and\ \citenamefont
  {Mena~Marug\'an}}]{Gomar2015}%
  \BibitemOpen
  \bibfield  {author} {\bibinfo {author} {\bibfnamefont {L.}~\bibnamefont
  {Castell\'o~Gomar}}, \bibinfo {author} {\bibfnamefont {M.}~\bibnamefont
  {Mart\'\i{}n-Benito}}, \ and\ \bibinfo {author} {\bibfnamefont {G.~A.}\
  \bibnamefont {Mena~Marug\'an}},\ }\href {\doibase
  10.1088/1475-7516/2015/06/045} {\bibfield  {journal} {\bibinfo  {journal}
  {JCAP}\ }\textbf {\bibinfo {volume} {06}},\ \bibinfo {pages} {045} (\bibinfo
  {year} {2015})},\ \Eprint {http://arxiv.org/abs/1503.03907} {arXiv:1503.03907
  [gr-qc]} \BibitemShut {NoStop}%
\bibitem [{\citenamefont {Elizaga~Navascu\'es}\ \emph
  {et~al.}(2018{\natexlab{a}})\citenamefont {Elizaga~Navascu\'es},
  \citenamefont {Martin~de Blas},\ and\ \citenamefont
  {Mena~Marug\'an}}]{hyb-vs-dress}%
  \BibitemOpen
  \bibfield  {author} {\bibinfo {author} {\bibfnamefont {B.}~\bibnamefont
  {Elizaga~Navascu\'es}}, \bibinfo {author} {\bibfnamefont {D.}~\bibnamefont
  {Martin~de Blas}}, \ and\ \bibinfo {author} {\bibfnamefont {G.~A.}\
  \bibnamefont {Mena~Marug\'an}},\ }\href {\doibase 10.1103/PhysRevD.97.043523}
  {\bibfield  {journal} {\bibinfo  {journal} {Phys. Rev. D}\ }\textbf {\bibinfo
  {volume} {97}},\ \bibinfo {pages} {043523} (\bibinfo {year}
  {2018}{\natexlab{a}})},\ \Eprint {http://arxiv.org/abs/1711.10861}
  {arXiv:1711.10861 [gr-qc]} \BibitemShut {NoStop}%
\bibitem [{\citenamefont {Elizaga~Navascu\'es}\ and\ \citenamefont
  {Mena~Marug\'an}(2021)}]{hybrid:ElizagaNavascues2021}%
  \BibitemOpen
  \bibfield  {author} {\bibinfo {author} {\bibfnamefont {B.}~\bibnamefont
  {Elizaga~Navascu\'es}}\ and\ \bibinfo {author} {\bibfnamefont {G.~A.}\
  \bibnamefont {Mena~Marug\'an}},\ }\href {\doibase
  10.1088/1475-7516/2021/09/030} {\bibfield  {journal} {\bibinfo  {journal}
  {JCAP}\ }\textbf {\bibinfo {volume} {09}},\ \bibinfo {pages} {030} (\bibinfo
  {year} {2021})},\ \Eprint {http://arxiv.org/abs/2104.15002} {arXiv:2104.15002
  [gr-qc]} \BibitemShut {NoStop}%
\bibitem [{\citenamefont {Iteanu}\ and\ \citenamefont
  {Mena~Marug\'an}(2022)}]{Iteanu2022}%
  \BibitemOpen
  \bibfield  {author} {\bibinfo {author} {\bibfnamefont {S.}~\bibnamefont
  {Iteanu}}\ and\ \bibinfo {author} {\bibfnamefont {G.~A.}\ \bibnamefont
  {Mena~Marug\'an}},\ }\href {\doibase 10.3390/universe8090463} {\bibfield
  {journal} {\bibinfo  {journal} {Universe}\ }\textbf {\bibinfo {volume} {8}},\
  \bibinfo {pages} {463} (\bibinfo {year} {2022})},\ \Eprint
  {http://arxiv.org/abs/2208.01987} {arXiv:2208.01987 [gr-qc]} \BibitemShut
  {NoStop}%
\bibitem [{\citenamefont {de~Blas}\ and\ \citenamefont
  {Olmedo}(2016)}]{deBlas2016}%
  \BibitemOpen
  \bibfield  {author} {\bibinfo {author} {\bibfnamefont {D.~M.}\ \bibnamefont
  {de~Blas}}\ and\ \bibinfo {author} {\bibfnamefont {J.}~\bibnamefont
  {Olmedo}},\ }\href {\doibase 10.1088/1475-7516/2016/06/029} {\bibfield
  {journal} {\bibinfo  {journal} {JCAP}\ }\textbf {\bibinfo {volume} {06}},\
  \bibinfo {pages} {029} (\bibinfo {year} {2016})},\ \Eprint
  {http://arxiv.org/abs/1601.01716} {arXiv:1601.01716 [gr-qc]} \BibitemShut
  {NoStop}%
\bibitem [{\citenamefont {Castell\'o~Gomar}\ \emph {et~al.}(2017)\citenamefont
  {Castell\'o~Gomar}, \citenamefont {Mena~Marug\'an}, \citenamefont
  {Mart\'\i{}n De~Blas},\ and\ \citenamefont {Olmedo}}]{CastelloGomar2017}%
  \BibitemOpen
  \bibfield  {author} {\bibinfo {author} {\bibfnamefont {L.}~\bibnamefont
  {Castell\'o~Gomar}}, \bibinfo {author} {\bibfnamefont {G.~A.}\ \bibnamefont
  {Mena~Marug\'an}}, \bibinfo {author} {\bibfnamefont {D.}~\bibnamefont
  {Mart\'\i{}n De~Blas}}, \ and\ \bibinfo {author} {\bibfnamefont
  {J.}~\bibnamefont {Olmedo}},\ }\href {\doibase 10.1103/PhysRevD.96.103528}
  {\bibfield  {journal} {\bibinfo  {journal} {Phys. Rev. D}\ }\textbf {\bibinfo
  {volume} {96}},\ \bibinfo {pages} {103528} (\bibinfo {year} {2017})},\
  \Eprint {http://arxiv.org/abs/1702.06036} {arXiv:1702.06036 [gr-qc]}
  \BibitemShut {NoStop}%
\bibitem [{\citenamefont {Elizaga~Navascu\'es}\ \emph
  {et~al.}(2018{\natexlab{b}})\citenamefont {Elizaga~Navascu\'es},
  \citenamefont {de~Blas},\ and\ \citenamefont
  {Marug\'an}}]{Elizaga_Navascu_s_2018}%
  \BibitemOpen
  \bibfield  {author} {\bibinfo {author} {\bibfnamefont {B.}~\bibnamefont
  {Elizaga~Navascu\'es}}, \bibinfo {author} {\bibfnamefont {D.~M.}\
  \bibnamefont {de~Blas}}, \ and\ \bibinfo {author} {\bibfnamefont {G.~A.~M.}\
  \bibnamefont {Marug\'an}},\ }\href {\doibase 10.3390/universe4100098}
  {\bibfield  {journal} {\bibinfo  {journal} {Universe}\ }\textbf {\bibinfo
  {volume} {4}},\ \bibinfo {pages} {98} (\bibinfo {year}
  {2018}{\natexlab{b}})},\ \Eprint {http://arxiv.org/abs/1809.09874}
  {arXiv:1809.09874 [gr-qc]} \BibitemShut {NoStop}%
\bibitem [{\citenamefont {Elizaga~Navascu\'es}\ \emph
  {et~al.}(2020)\citenamefont {Elizaga~Navascu\'es}, \citenamefont
  {Marug\'an},\ and\ \citenamefont {Prado}}]{menava}%
  \BibitemOpen
  \bibfield  {author} {\bibinfo {author} {\bibfnamefont {B.}~\bibnamefont
  {Elizaga~Navascu\'es}}, \bibinfo {author} {\bibfnamefont {G.~A.~M.}\
  \bibnamefont {Marug\'an}}, \ and\ \bibinfo {author} {\bibfnamefont
  {S.}~\bibnamefont {Prado}},\ }\href {\doibase 10.1088/1361-6382/abc6bb}
  {\bibfield  {journal} {\bibinfo  {journal} {Class. Quant. Grav.}\ }\textbf
  {\bibinfo {volume} {38}},\ \bibinfo {pages} {035001} (\bibinfo {year}
  {2020})},\ \Eprint {http://arxiv.org/abs/2005.10194} {arXiv:2005.10194
  [gr-qc]} \BibitemShut {NoStop}%
\bibitem [{\citenamefont {Elizaga~Navascu\'es}\ \emph
  {et~al.}(2021)\citenamefont {Elizaga~Navascu\'es}, \citenamefont
  {Mena~Marug\'an},\ and\ \citenamefont {Prado}}]{ElizagaNavascues2021}%
  \BibitemOpen
  \bibfield  {author} {\bibinfo {author} {\bibfnamefont {B.}~\bibnamefont
  {Elizaga~Navascu\'es}}, \bibinfo {author} {\bibfnamefont {G.~A.}\
  \bibnamefont {Mena~Marug\'an}}, \ and\ \bibinfo {author} {\bibfnamefont
  {S.}~\bibnamefont {Prado}},\ }\href {\doibase 10.1103/PhysRevD.104.083541}
  {\bibfield  {journal} {\bibinfo  {journal} {Phys. Rev. D}\ }\textbf {\bibinfo
  {volume} {104}},\ \bibinfo {pages} {083541} (\bibinfo {year} {2021})},\
  \Eprint {http://arxiv.org/abs/2107.11054} {arXiv:2107.11054 [gr-qc]}
  \BibitemShut {NoStop}%
\bibitem [{\citenamefont {Mart\'\i{}n-Benito}\ \emph
  {et~al.}(2021{\natexlab{a}})\citenamefont {Mart\'\i{}n-Benito}, \citenamefont
  {Neves},\ and\ \citenamefont {Olmedo}}]{mno-sles}%
  \BibitemOpen
  \bibfield  {author} {\bibinfo {author} {\bibfnamefont {M.}~\bibnamefont
  {Mart\'\i{}n-Benito}}, \bibinfo {author} {\bibfnamefont {R.~B.}\ \bibnamefont
  {Neves}}, \ and\ \bibinfo {author} {\bibfnamefont {J.}~\bibnamefont
  {Olmedo}},\ }\href {\doibase 10.3389/fspas.2021.702543} {\bibfield  {journal}
  {\bibinfo  {journal} {Front. Astron. Space Sci.}\ }\textbf {\bibinfo {volume}
  {0}},\ \bibinfo {pages} {133} (\bibinfo {year} {2021}{\natexlab{a}})},\
  \Eprint {http://arxiv.org/abs/2104.14850} {arXiv:2104.14850 [gr-qc]}
  \BibitemShut {NoStop}%
\bibitem [{\citenamefont {Mart\'\i{}n-Benito}\ \emph
  {et~al.}(2021{\natexlab{b}})\citenamefont {Mart\'\i{}n-Benito}, \citenamefont
  {Neves},\ and\ \citenamefont {Olmedo}}]{Martin-Benito2021}%
  \BibitemOpen
  \bibfield  {author} {\bibinfo {author} {\bibfnamefont {M.}~\bibnamefont
  {Mart\'\i{}n-Benito}}, \bibinfo {author} {\bibfnamefont {R.~B.}\ \bibnamefont
  {Neves}}, \ and\ \bibinfo {author} {\bibfnamefont {J.}~\bibnamefont
  {Olmedo}},\ }\href {\doibase 10.1103/PhysRevD.103.123524} {\bibfield
  {journal} {\bibinfo  {journal} {Phys. Rev. D}\ }\textbf {\bibinfo {volume}
  {103}},\ \bibinfo {pages} {123524} (\bibinfo {year} {2021}{\natexlab{b}})},\
  \Eprint {http://arxiv.org/abs/2104.03035} {arXiv:2104.03035 [gr-qc]}
  \BibitemShut {NoStop}%
\bibitem [{\citenamefont {Fixsen}\ \emph {et~al.}(1996)\citenamefont {Fixsen},
  \citenamefont {Cheng}, \citenamefont {Gales}, \citenamefont {Mather},
  \citenamefont {Shafer},\ and\ \citenamefont {Wright}}]{Fixsen:1996nj}%
  \BibitemOpen
  \bibfield  {author} {\bibinfo {author} {\bibfnamefont {D.~J.}\ \bibnamefont
  {Fixsen}}, \bibinfo {author} {\bibfnamefont {E.~S.}\ \bibnamefont {Cheng}},
  \bibinfo {author} {\bibfnamefont {J.~M.}\ \bibnamefont {Gales}}, \bibinfo
  {author} {\bibfnamefont {J.~C.}\ \bibnamefont {Mather}}, \bibinfo {author}
  {\bibfnamefont {R.~A.}\ \bibnamefont {Shafer}}, \ and\ \bibinfo {author}
  {\bibfnamefont {E.~L.}\ \bibnamefont {Wright}},\ }\href {\doibase
  10.1086/178173} {\bibfield  {journal} {\bibinfo  {journal} {Astrophys. J.}\
  }\textbf {\bibinfo {volume} {473}},\ \bibinfo {pages} {576} (\bibinfo {year}
  {1996})},\ \Eprint {http://arxiv.org/abs/astro-ph/9605054}
  {arXiv:astro-ph/9605054} \BibitemShut {NoStop}%
\bibitem [{\citenamefont {Di~Valentino}\ \emph {et~al.}(2019)\citenamefont
  {Di~Valentino}, \citenamefont {Melchiorri},\ and\ \citenamefont
  {Silk}}]{DiValentino2019}%
  \BibitemOpen
  \bibfield  {author} {\bibinfo {author} {\bibfnamefont {E.}~\bibnamefont
  {Di~Valentino}}, \bibinfo {author} {\bibfnamefont {A.}~\bibnamefont
  {Melchiorri}}, \ and\ \bibinfo {author} {\bibfnamefont {J.}~\bibnamefont
  {Silk}},\ }\href {\doibase 10.1038/s41550-019-0906-9} {\bibfield  {journal}
  {\bibinfo  {journal} {Nature Astron.}\ }\textbf {\bibinfo {volume} {4}},\
  \bibinfo {pages} {196} (\bibinfo {year} {2019})},\ \Eprint
  {http://arxiv.org/abs/1911.02087} {arXiv:1911.02087 [astro-ph.CO]}
  \BibitemShut {NoStop}%
\bibitem [{\citenamefont {Blas}\ \emph {et~al.}(2011)\citenamefont {Blas},
  \citenamefont {Lesgourgues},\ and\ \citenamefont {Tram}}]{Blas_2011}%
  \BibitemOpen
  \bibfield  {author} {\bibinfo {author} {\bibfnamefont {D.}~\bibnamefont
  {Blas}}, \bibinfo {author} {\bibfnamefont {J.}~\bibnamefont {Lesgourgues}}, \
  and\ \bibinfo {author} {\bibfnamefont {T.}~\bibnamefont {Tram}},\ }\href
  {\doibase 10.1088/1475-7516/2011/07/034} {\bibfield  {journal} {\bibinfo
  {journal} {Journal of Cosmology and Astroparticle Physics}\ }\textbf
  {\bibinfo {volume} {2011}},\ \bibinfo {pages} {034} (\bibinfo {year}
  {2011})},\ \Eprint {http://arxiv.org/abs/1104.2933} {1104.2933} \BibitemShut
  {NoStop}%
\bibitem [{\citenamefont {Brinckmann}\ and\ \citenamefont
  {Lesgourgues}(2019)}]{Brinckmann2018}%
  \BibitemOpen
  \bibfield  {author} {\bibinfo {author} {\bibfnamefont {T.}~\bibnamefont
  {Brinckmann}}\ and\ \bibinfo {author} {\bibfnamefont {J.}~\bibnamefont
  {Lesgourgues}},\ }\href {\doibase 10.1016/j.dark.2018.100260} {\bibfield
  {journal} {\bibinfo  {journal} {Phys. Dark Univ.}\ }\textbf {\bibinfo
  {volume} {24}},\ \bibinfo {pages} {100260} (\bibinfo {year} {2019})},\
  \Eprint {http://arxiv.org/abs/1804.07261} {arXiv:1804.07261 [astro-ph.CO]}
  \BibitemShut {NoStop}%
\bibitem [{\citenamefont {Audren}\ \emph {et~al.}(2013)\citenamefont {Audren},
  \citenamefont {Lesgourgues}, \citenamefont {Benabed},\ and\ \citenamefont
  {Prunet}}]{Audren2012}%
  \BibitemOpen
  \bibfield  {author} {\bibinfo {author} {\bibfnamefont {B.}~\bibnamefont
  {Audren}}, \bibinfo {author} {\bibfnamefont {J.}~\bibnamefont {Lesgourgues}},
  \bibinfo {author} {\bibfnamefont {K.}~\bibnamefont {Benabed}}, \ and\
  \bibinfo {author} {\bibfnamefont {S.}~\bibnamefont {Prunet}},\ }\href
  {\doibase 10.1088/1475-7516/2013/02/001} {\bibfield  {journal} {\bibinfo
  {journal} {JCAP}\ }\textbf {\bibinfo {volume} {1302}},\ \bibinfo {pages}
  {001} (\bibinfo {year} {2013})},\ \Eprint {http://arxiv.org/abs/1210.7183}
  {arXiv:1210.7183 [astro-ph.CO]} \BibitemShut {NoStop}%
\bibitem [{\citenamefont {Aghanim}\ \emph
  {et~al.}(2020{\natexlab{a}})\citenamefont {Aghanim} \emph
  {et~al.}}]{PlanckV}%
  \BibitemOpen
  \bibfield  {author} {\bibinfo {author} {\bibfnamefont {N.}~\bibnamefont
  {Aghanim}} \emph {et~al.} (\bibinfo {collaboration} {Planck}),\ }\href
  {\doibase 10.1051/0004-6361/201936386} {\bibfield  {journal} {\bibinfo
  {journal} {Astron. Astrophys.}\ }\textbf {\bibinfo {volume} {641}},\ \bibinfo
  {pages} {A5} (\bibinfo {year} {2020}{\natexlab{a}})},\ \Eprint
  {http://arxiv.org/abs/1907.12875} {arXiv:1907.12875 [astro-ph.CO]}
  \BibitemShut {NoStop}%
\bibitem [{\citenamefont {Aghanim}\ \emph
  {et~al.}(2020{\natexlab{b}})\citenamefont {Aghanim} \emph
  {et~al.}}]{PlanckVIII}%
  \BibitemOpen
  \bibfield  {author} {\bibinfo {author} {\bibfnamefont {N.}~\bibnamefont
  {Aghanim}} \emph {et~al.} (\bibinfo {collaboration} {Planck}),\ }\href
  {\doibase 10.1051/0004-6361/201833886} {\bibfield  {journal} {\bibinfo
  {journal} {Astron. Astrophys.}\ }\textbf {\bibinfo {volume} {641}},\ \bibinfo
  {pages} {A8} (\bibinfo {year} {2020}{\natexlab{b}})},\ \Eprint
  {http://arxiv.org/abs/1807.06210} {arXiv:1807.06210 [astro-ph.CO]}
  \BibitemShut {NoStop}%
\bibitem [{\citenamefont {Contaldi}\ \emph {et~al.}(2003)\citenamefont
  {Contaldi}, \citenamefont {Peloso}, \citenamefont {Kofman},\ and\
  \citenamefont {Linde}}]{Contaldi2003}%
  \BibitemOpen
  \bibfield  {author} {\bibinfo {author} {\bibfnamefont {C.~R.}\ \bibnamefont
  {Contaldi}}, \bibinfo {author} {\bibfnamefont {M.}~\bibnamefont {Peloso}},
  \bibinfo {author} {\bibfnamefont {L.}~\bibnamefont {Kofman}}, \ and\ \bibinfo
  {author} {\bibfnamefont {A.~D.}\ \bibnamefont {Linde}},\ }\href {\doibase
  10.1088/1475-7516/2003/07/002} {\bibfield  {journal} {\bibinfo  {journal}
  {JCAP}\ }\textbf {\bibinfo {volume} {07}},\ \bibinfo {pages} {002} (\bibinfo
  {year} {2003})},\ \Eprint {http://arxiv.org/abs/astro-ph/0303636}
  {arXiv:astro-ph/0303636} \BibitemShut {NoStop}%
\bibitem [{\citenamefont {Gupt}(2017)}]{Gupt2017}%
  \BibitemOpen
  \bibfield  {author} {\bibinfo {author} {\bibfnamefont {B.}~\bibnamefont
  {Gupt}},\ }\href@noop {} {\  (\bibinfo {year} {2017})},\ \Eprint
  {http://arxiv.org/abs/1710.00759} {arXiv:1710.00759 [astro-ph.CO]}
  \BibitemShut {NoStop}%
\bibitem [{\citenamefont {Taveras}(2008)}]{Taveras:2008ke}%
  \BibitemOpen
  \bibfield  {author} {\bibinfo {author} {\bibfnamefont {V.}~\bibnamefont
  {Taveras}},\ }\href {\doibase 10.1103/PhysRevD.78.064072} {\bibfield
  {journal} {\bibinfo  {journal} {Phys. Rev. D}\ }\textbf {\bibinfo {volume}
  {78}},\ \bibinfo {pages} {064072} (\bibinfo {year} {2008})},\ \Eprint
  {http://arxiv.org/abs/0807.3325} {arXiv:0807.3325 [gr-qc]} \BibitemShut
  {NoStop}%
\bibitem [{\citenamefont {Diener}\ \emph {et~al.}(2014)\citenamefont {Diener},
  \citenamefont {Gupt},\ and\ \citenamefont {Singh}}]{Diener:2014mia}%
  \BibitemOpen
  \bibfield  {author} {\bibinfo {author} {\bibfnamefont {P.}~\bibnamefont
  {Diener}}, \bibinfo {author} {\bibfnamefont {B.}~\bibnamefont {Gupt}}, \ and\
  \bibinfo {author} {\bibfnamefont {P.}~\bibnamefont {Singh}},\ }\href
  {\doibase 10.1088/0264-9381/31/10/105015} {\bibfield  {journal} {\bibinfo
  {journal} {Class. Quant. Grav.}\ }\textbf {\bibinfo {volume} {31}},\ \bibinfo
  {pages} {105015} (\bibinfo {year} {2014})},\ \Eprint
  {http://arxiv.org/abs/1402.6613} {arXiv:1402.6613 [gr-qc]} \BibitemShut
  {NoStop}%
\bibitem [{\citenamefont {Bonga}\ and\ \citenamefont {Gupt}(2016)}]{Bonga2015}%
  \BibitemOpen
  \bibfield  {author} {\bibinfo {author} {\bibfnamefont {B.}~\bibnamefont
  {Bonga}}\ and\ \bibinfo {author} {\bibfnamefont {B.}~\bibnamefont {Gupt}},\
  }\href {\doibase 10.1007/s10714-016-2071-0} {\bibfield  {journal} {\bibinfo
  {journal} {Gen. Rel. Grav.}\ }\textbf {\bibinfo {volume} {48}},\ \bibinfo
  {pages} {71} (\bibinfo {year} {2016})},\ \Eprint
  {http://arxiv.org/abs/1510.00680} {arXiv:1510.00680 [gr-qc]} \BibitemShut
  {NoStop}%
\end{thebibliography}%

\end{document}